\documentclass[prb,aps,epsf,twocolumn,showpacs,10pt]{revtex4-1}
\usepackage{graphicx,amsfonts,times,bm,amsmath,verbatim,color,array}

\begin{document}
\title{Interplay of valley polarization and dynamic nuclear polarization in 2D transition metal dichalcogenides }

\author{Girish Sharma}
\author{Sophia E. Economou}
\author{Edwin Barnes}

\affiliation{Department of Physics, Virginia Tech, Blacksburg, VA 24061, U.S.A}

\begin{abstract}
The interplay of Ising spin-orbit coupling and non-trivial band topology in transition metal dichalcogenides (TMDs) produces anomalous transport and optical properties that are very different from a regular 2D electron gas. The spin-momentum locking of optically excited carriers near a valley point can give rise to an anomalous spin-valley Hall current under the application of an in-plane electric field. TMDs also exhibit strong electron-nuclear hyperfine interactions, but their effect on spin-valley-locked currents remains unknown. Here, we show that hyperfine interactions can create a feedback mechanism in which spin-valley currents generate significant dynamical nuclear polarization which in turn Zeeman shifts excitonic transitions out of resonance with an optical driving field, saturating the production of spin-valley polarization. We propose an experimental signature of dynamic nuclear polarization which can be detected via measurements of the anomalous Hall current. Our results help to elucidate the interplay of valley polarization and nuclear spin dynamics in TMDs.

\end{abstract}

\maketitle

\section{Introduction}
Atomically thin 2D van der Waals-bonded materials offer new possibilities both from a fundamental physics perspective and in terms of potential technological applications~\cite{Geim:2013, Novoselov:2005, Gomez:2016}. Among these, semiconducting transition metal dichalcogenides (TMDs) of the form MX$_2$, where M$=$Mo, W, and X$=$S, Se have received special attention because of the possibility of manipulating the valley degree of freedom~\cite{Xiao:2012, Xu:2014, Yang:2015, Hao:2016, Mak:2014, Mak:2016} in addition to spin and charge. This feature is highly promising for low-power electronics and valley-tronics applications such as faster computer logic systems and data storage chips in next-generation devices. In TMDs, a broken in-plane mirror symmetry results in a special intrinsic spin-orbit coupling (SOC) called Ising SOC~\cite{{Zhu:2011}, {Kormanyos:2013}, {Zahid:2013}, {Cappelluti:2013}}, which acts as an effective valley-dependent Zeeman field, strongly polarizing the electron spins perpendicular to the 2D plane, in sharp contrast to the 2D helical liquid produced by the more familiar Rashba SOC~\cite{Manchon:2015}. On account of this Zeeman-type spin-splitting, the valley bands in TMDs are automatically spin-locked even though overall time-reversal symmetry is preserved---a distinctive property of these materials. These striking features have generated enormous interest in TMDs and spurred substantial experimental progress over the past few years~\cite{Mak:2016, Xu:2014, Mak:2014, Yang:2015, Hao:2016, Lee:2016, Wu:2013, {Zhu:2011}, {Kormanyos:2013}, {Zahid:2013}, {Cappelluti:2013}, Jonker1, Jonker2, Jonker3, Jonker4, Jonker5, XXu1, XXu2, XXu3, XXu4}.

Electrons in a semiconductor are subject to various spin-dependent interactions with the environment which can cause relaxation and decoherence, such as coupling to phonons, magnetic and non-magnetic impurities, nuclear spins, and carrier-carrier scattering. Out of the many sources of decoherence, electron-nuclear spin interactions play a prominent role~\cite{Yao:2006, Cywinski:2009, Liu:2010, Barnes:2012, Coish:2010, Koppens:2006, Kloeffel:2011} both for electrons confined in a quantum dot and also for optically excited carriers, especially at low temperatures when other mechanisms can be suppressed. In TMDs, both the transition element M and the chalcogen X have stable elemental isotopes with nonzero nuclear spin, and therefore one expects interesting electron-nuclear spin dynamics to occur in these materials. Apart from the nuclear spins acting as a decoherence channel, the possibility of optical control of the electron-nuclear spin entanglement has also been pointed out recently in TMD-based quantum dot systems~\cite{WYao:2016}. However, the role of dynamic nuclear polarization (DNP) in TMDs has not been investigated so far, despite its striking effects in other low dimensional semiconductor systems~\cite{Greilich:2007,Vink:2009,Latta:2009,  XXu:2009,Foletti:2009,Carter:2009,Kou:2010,Hogele:2012,Carter:2014, Tian:2017,Burkard:2009}.

\begin{figure}
\includegraphics[scale=0.3]{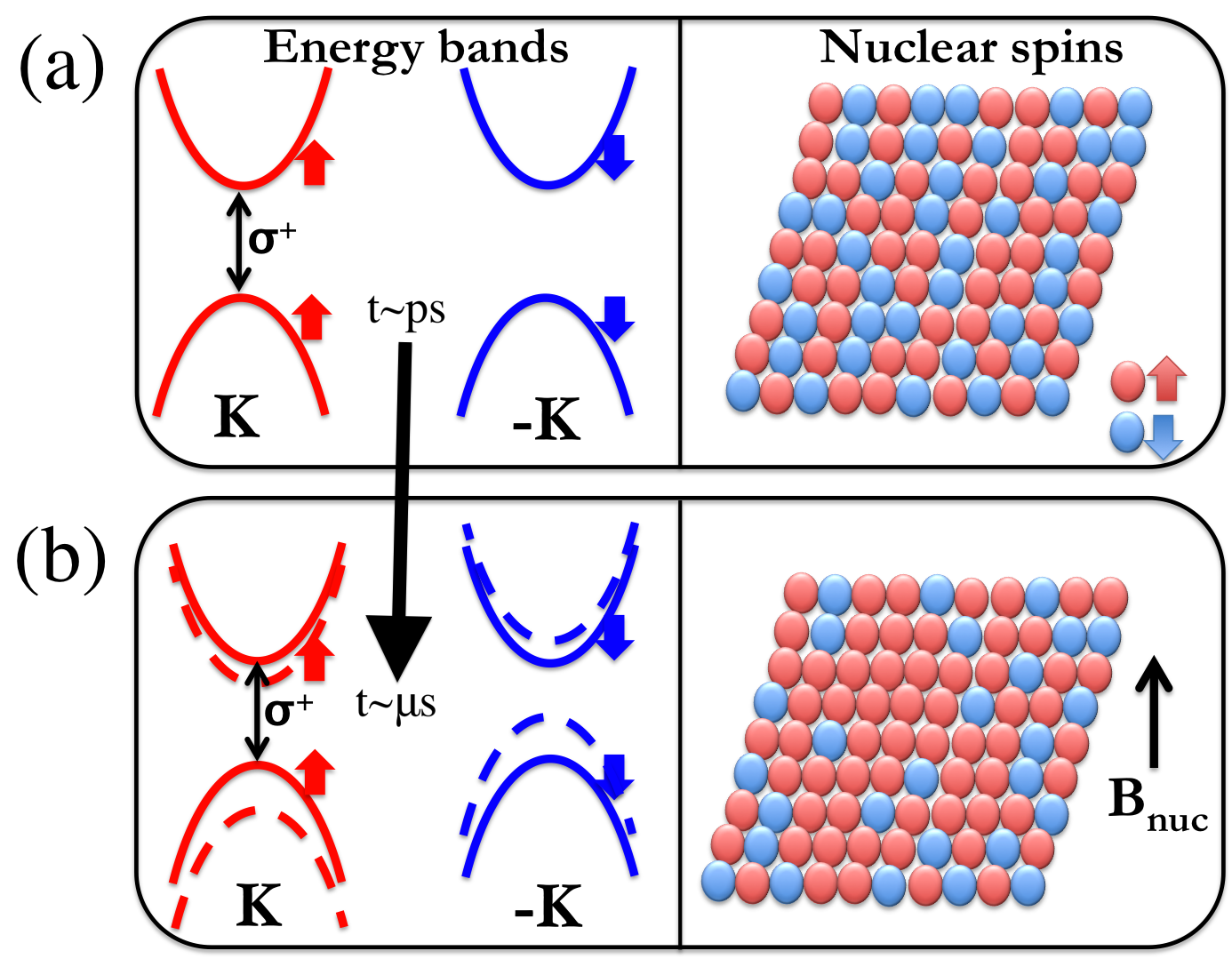}
\caption{(Color online) Schematic illustration of the feedback effect via hyperfine interaction in TMDs. Left panel indicates the energy bands near the valley points, and the right panel indicates the nuclear spins. \textit{(a)} Excitation of $\mathbf{K}$ valley via $\sigma^+$ CP light produces spin-polarized carriers on a $ps$ timescale. The $-\mathbf{K}$ valley is not excited, and the underlying nuclear spins are unpolarized. The arrows on the left (right) indicate the polarization of electron (nuclear) spins. \textit{(b)} Hyperfine interaction causes nuclear spins to polarize, however at a much slower time scale. The effective nuclear field $B_{\text{nuc}}$ shifts the energy spectrum (by different magnitudes in conduction/valence bands), thus detuning the laser and damping the carrier population. This effect can be measured in transport experiments such as the anomalous Hall current and provides a clear signature of nuclear spins.}
\label{Fig_1}
\end{figure}

In this Letter, we show that optically-pumped spin-valley currents in TMDs can create significant dynamic nuclear polarization (DNP), which in turn has important consequences for spin-valley polarization and transport in these materials. A feedback mechanism arises in which DNP causes a Zeeman shift of the optically driven valence and conduction bands, forcing the system out of resonance with the driving field and saturating the production of spin-valley polarization, as illustrated in Fig.~\ref{Fig_1}. We perform a self-consistent calculation of the anomalous spin-valley Hall current and consequent DNP buildup and show that feedback between the two leads to clear signatures that can be detected via transport measurements. Our results can serve as a prototype for DNP-based experiments in TMDs, uncovering the interplay of valley polarization and nuclear spin dynamics.

Our calculation proceeds as follows. We first analytically solve for the coupled dynamics of the electron and hole distribution functions for a TMD subject to non-perturbative optical driving using the formalism of semiconductor Bloch equations (SBEs)~\cite{Haug:2009}. 
We then incorporate electron-nuclear hyperfine dynamics by solving a Lindblad equation, using the solution to the SBEs as the initial condition. This approach takes advantage of the fact that the hyperfine dynamics are much slower than the timescales of optical excitation and carrier transport. We extract effective nuclear spin flip rates from the Lindblad equation and input these into a kinetic equation for the nuclear polarization, the solution of which gives the steady state DNP. Finally, this steady state value is fed back into the SBE solution to obtain the effect of DNP on the carrier distribution functions and anomalous Hall current. Our formalism thus provides a self-consistent description of the rich feedback mechanism arising from optical driving, spin-valley locking, and hyperfine interactions in TMDs.

\section{Carrier dynamics in TMDs}
The effective Hamiltonian for a generic monolayer TMD near a valley point ($\mathbf{K}$ point) can be written as~\cite{Xiao:2012}
\begin{eqnarray}
H_{\kappa,s,\mathbf{k}} = at' (\kappa k_x \tau_x {+} k_y \tau_y)+ \frac{\Delta}{2}\tau_z - \frac{\tau_z {-} 1}{2}\lambda\kappa s,\nonumber\\
\label{Eq_H0}
\end{eqnarray}
where $a$ is the lattice constant, $t'$ the effective hopping integral, $\Delta$ is the energy gap, and $2\lambda$ is the spin splitting for the valence band caused by SOC. The binary indices $\kappa=\pm1$ and $s=\pm1$ correspond to valley and spin respectively, and $\tau$ are Pauli matrices in the conduction-valence band subspace. Due to the valence band spin-splitting, the effective band-gap becomes $\Delta'=\Delta-\kappa s\lambda$. In the above equation we have neglected Coulomb interactions since their primary effect is to renormalize single-particle energies and the Rabi frequency. We also neglect the conduction band spin-splitting since this is much smaller than the valence spin-splitting. We will confirm later on that inclusion of the conduction spin-splitting does not produce qualitative changes in nuclear feedback effects.
The inter-band coupling to the light field can be described by the following Hamiltonian in the $\tau$ basis:
\begin{eqnarray}
H_I = -\mathcal{E}(t) \begin{pmatrix}
  0 & d_{\mathbf{k}} \\
  d^*_{\mathbf{k}} & 0
 \end{pmatrix},
 \label{Eq_H_I}
\end{eqnarray}
where the matrix element $d_\mathbf{k}$ for $\sigma^+$ light is~\cite{Xiao:2012}
\begin{eqnarray}
d_{\mathbf{k}} = ie\frac{at'}{\omega_{cv\mathbf{k}}}\left(1+\kappa\frac{\Delta'}{\sqrt{\Delta'^2 + 4a^2t'^2k^2}}\right) 
\end{eqnarray}
For a given valley index, $d_{\mathbf{K}}\sim (1+\kappa)$, which indicates a selective valley coupling  to CP light. In the above equation $\omega_{cv\mathbf{k}}$ is the energy splitting between the two eigenvalues of $H_{\kappa,s,\mathbf{k}}$.

\begin{figure}
\includegraphics[scale=0.215]{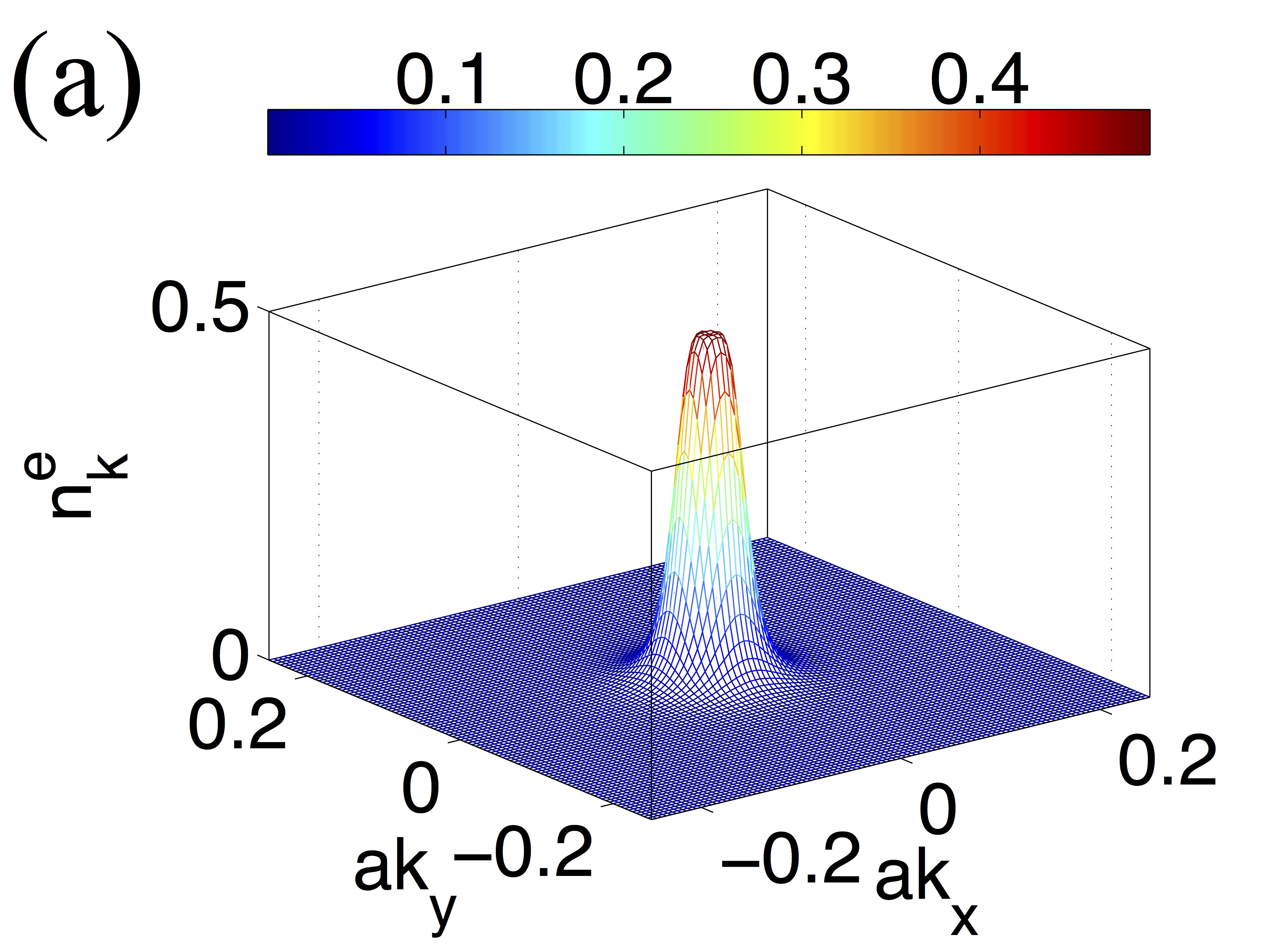}
\includegraphics[scale=0.215]{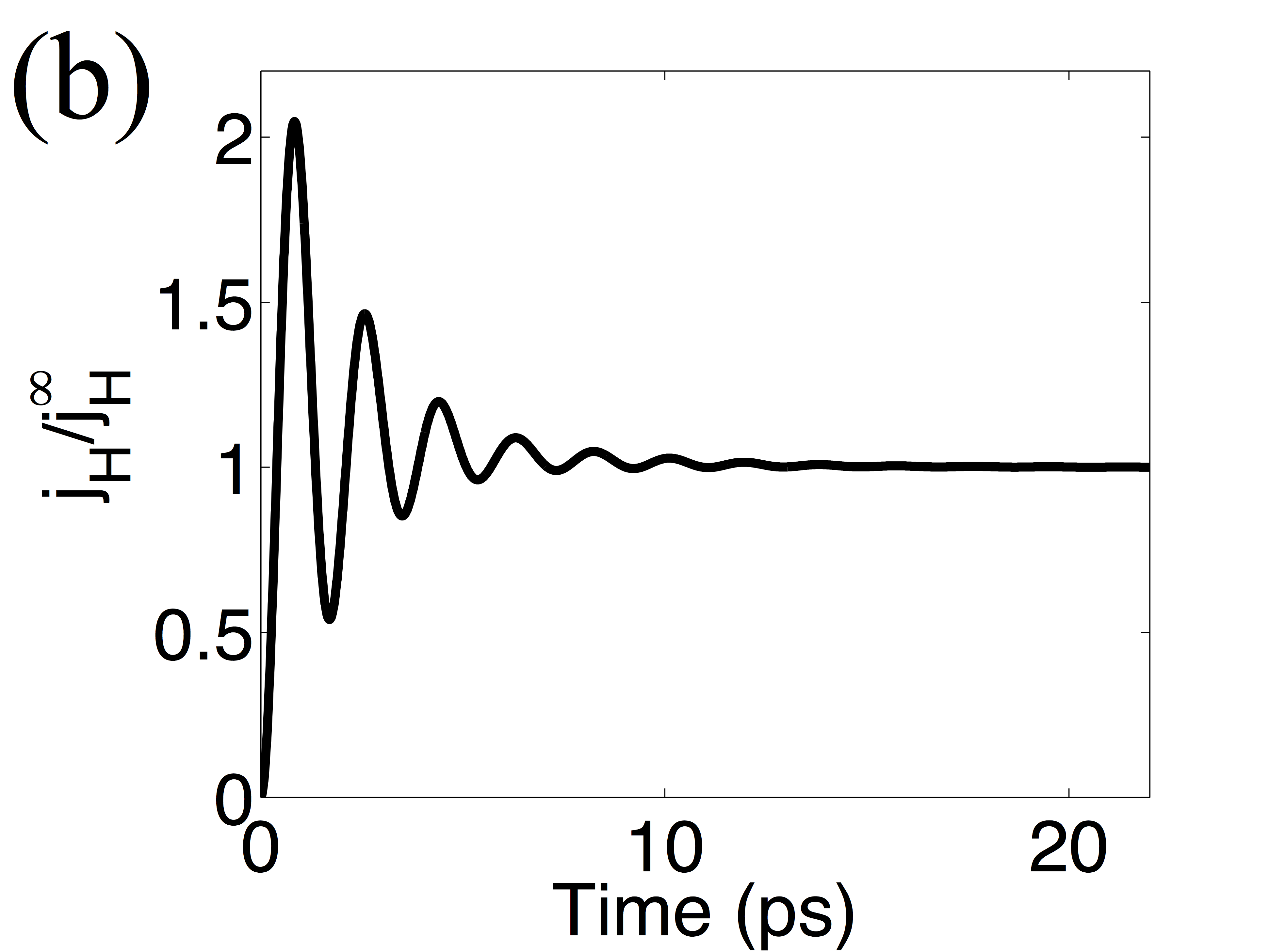}
\caption{(Color online) \textit{(a)} Steady state electron distribution function $n^{\infty}_{e\mathbf{k}}$ in the $\mathbf{K}$ valley, as obtained from the solution to  Eqs.~\eqref{Eq_Pk_1}-\eqref{Eq_nhk_1} for $\sigma^+$ optical excitation. The distribution function is sharply peaked around the $\mathbf{K}$ point. \textit{(b)} Hall current (normalized with respect to the steady state value $j_H^\infty$) obtained from Eq.~\eqref{Eq_J_HAll}. The Hall current reaches a steady state on a $ps$ time scale for typical material parameters~\cite{Mai, QWang:2013} $\gamma=5ps$, $\Gamma=10ps$, and $\Omega_{R\mathbf{K}}=1meV$. }
\label{Fig_2}
\end{figure}

Though one can treat the problem of light-matter interaction even from a time-dependent quasiclassical Boltzmann approach (see Appendix A), it proves advantageous to obtain an exact quantum mechanical solution. Our goal is to calculate the laser-induced dynamics of the following quantities: $n_{e\mathbf{k}}=\langle c^\dagger_{c,\mathbf{k}} c_{c,\mathbf{k}}\rangle$, $n_{h\mathbf{k}}=\langle c_{v,\mathbf{k}} c^\dagger_{v,\mathbf{k}}\rangle$, $Q_{\mathbf{k}}=\langle c^\dagger_{v\mathbf{k}}c_{c\mathbf{k}}\rangle$, which denote the conduction and valence band distribution functions, and the inter-band polarization, respectively. The Heisenberg equations of motion for the total Hamiltonian, $H=H_{\mathbf{k}}+H_I$, reduce to the following set of equations describing laser induced carrier dynamics~\cite{Haug:2009}
\begin{align}
\frac{d Q_{\mathbf{k}}}{dt} &= -i(e_{e\mathbf{k}}+e_{h\mathbf{k}})Q_{\mathbf{k}} - i(n_{e\mathbf{k}}+n_{h\mathbf{k}}-1)\Omega_{R\mathbf{k}} \nonumber \\
&+ \left[\frac{d Q_{\mathbf{k}}}{dt}\right]_{\text{scatt}}, \label{Eq_Pk_1}\\
\frac{d n_{e\mathbf{k}}}{dt} &= -2 \text{Im} (\Omega_{R\mathbf{k}}Q_{\mathbf{k}}^*)+ \left[\frac{d n_{e\mathbf{k}}}{dt}\right]_{\text{scatt}},\label{Eq_nek_1}\\
\frac{d n_{h\mathbf{k}}}{dt} &= -2 \text{Im} (\Omega_{R\mathbf{k}}Q_{\mathbf{k}}^*)+ \left[\frac{d n_{h\mathbf{k}}}{dt}\right]_{\text{scatt}}, \label{Eq_nhk_1}
\end{align}
where $e_{e(h)\mathbf{k}}$ describe the electron(hole) renormalized single-particle energies, and $\Omega_{R\mathbf{k}}$ is the Rabi frequency. In the above equations, the scattering terms $[d\langle A\rangle/dt]_{\text{scatt}}$ on the right denote the difference between the full derivatives and the Hartree-Fock terms. 
We introduce two timescales: $\gamma$ for inter-band depolarization, and $\Gamma$ for intra-band carrier scattering. Since the inter-valley scattering rate is much slower than $\Gamma$, we can safely ignore such processes here. For $\Gamma\ll\gamma$, even a self-consistent solution to the Boltzmann transport equation (see Appendix A) can effectively describe band populations, but the present approach works for the general case.
The SBEs (Eq.~\eqref{Eq_Pk_1}-\eqref{Eq_nhk_1}) yield the following analytical solution (see Appendix B)
\begin{align}
w_{\mathbf{k}}(t) =1-2n_{e\mathbf{k}}&= A_1e^{-\alpha_1t} + A_2e^{-\alpha_2t} \cos(\alpha_3t) \nonumber \\
&+ (A_3/\alpha_3) e^{-\alpha_2t} \sin(\alpha_3t) + A_4.
\label{Eq_w_t_1}
\end{align}
The $\mathbf{k}-$dependent functions $\{A_i, \alpha_i\}$ are defined in Appendix B. The electron distribution function reaches a steady state given by {$n_{e\mathbf{k}}^{\infty} = (1-A_4)/2$}. This result is shown in Fig.~\ref{Fig_2} for the case of $\sigma^+$ optical driving. We see that the electron population is sharply peaked around the $\mathbf{K}$ valley point.

\section{Anomalous Hall current} Before we incorporate nuclear effects into our formalism, we first consider the anomalous Hall current in the absence of hyperfine interactions for later comparison. The intrinsic contribution to the anomalous Hall conductivity is linked to the topological properties of the Bloch states and essentially reduces to the integral of Berry phases over cuts of Fermi surface segments~\cite{Haldane:2004}. A non-trivial Berry phase associated with Bloch electrons gives the valley Hall and spin Hall effects in TMDs~\cite{Xiao:2012}. The Berry curvature for a Bloch wavefunction $|u_{n\mathbf{k}}\rangle$ is defined as $\Omega_{n\mathbf{k}}=\hat{z}\cdot \nabla_\mathbf{k}\times\langle u_{n\mathbf{k}}|i\nabla_{\mathbf{k}}|u_{n\mathbf{k}}\rangle$. The Berry curvature for the effective Hamiltonian defined in Eq.~\eqref{Eq_H0} is
\begin{eqnarray}
\Omega_n(\mathbf{k})=\kappa(-1)^n\frac{2a^2t'^2\Delta'^2}{(\Delta'^2 + 4a^2t'^2k^2)^{3/2}},
\label{Eq_Berry1}
\end{eqnarray}
where $n$ is the band index. 
In two dimensions, however, the Berry phase modified equations become $\mathbf{\dot{r}}_n =  \mathbf{v}_{n\mathbf{k}} + \frac{e}{\hbar} (\mathbf{E}\times\mathbf{\Omega}_{n\mathbf{k}})$ and $\hbar \mathbf{\dot{k}}_n =  -e\mathbf{E}$ in the absence of any magnetic field. Thus, in the presence of an in-plane electric field, electrons acquire an anomalous transverse velocity proportional to the Berry curvature~\cite{Niu:2010}. Using the electron distribution function $n_{e\mathbf{k}}$ obtained earlier, the Hall current can be calculated from
\begin{eqnarray}
\mathbf{j_{\text{Hall}}} = \sum\limits_\kappa\sum\limits_i\int{[d\mathbf{k}] \mathbf{\dot{r}}_{i\kappa} n_{i\mathbf{k}\kappa}},
\label{Eq_J_HAll}
\end{eqnarray}
where we take into account both valleys indexed by $\kappa$, and $i$ is the electron/hole index. Fig.~\ref{Fig_2} shows the time evolution of the Hall current for $\sigma^+$ optical excitation. For typical parameter values, the Hall current reaches a steady state on picosecond timescales.

\section{Coupling with nuclei (DNP)} In TMDs, both the transition element M and the chalcogen X have stable isotopes with nonzero nuclear spin, and therefore one expects interesting electron-nuclear spin dynamics. For our calculations, we include only the hyperfine interaction with the $M$ atom, which is dominant, based on the combination of natural abundance and hyperfine strength~\cite{WYao:2016}. Going beyond this approximation would only result in small quantitative corrections. We also take advantage of the timescale hierarchy, i.e. the time it takes for the electron distribution function to reach its steady state is much smaller than the characteristic timescale for nuclear dynamics (see Appendix C).
Ignoring inter-valley contributions, the hyperfine interaction near a valley point is \cite{WYao:2016}
\begin{align}
H_{\text{hyp}}&=\frac{A^e}{2} \sigma_z s_z (\tau_0+\tau_z) + \frac{A^h}{2} \sigma_z s_z (\tau_0-\tau_z)\nonumber\\
&+ \frac{\eta A^e}{4} (\sigma_x s_x +\sigma_y s_y)(\tau_0+\tau_z),
\end{align}
where $A^e$ and $A^h$ are the Overhauser hyperfine couplings for electron and holes respectively, and $\eta A^e$ gives the magnitude of the spin flip-flop term in the conduction band. The Pauli matrices $\sigma_i$ correspond to the nuclear spin. Note that due to the large spin-orbit splitting, the flip-flop term in the hole band is suppressed.

We first consider coupling the electron and hole to a single nuclear spin. Because the electrons and holes are continuously replenished through recombination and optical pumping, one might expect that they are subject to decoherence at a rate comparable to the recombination rate $\gamma$. However, because the hyperfine flip-flop interaction causes the electron spin to rotate relative to the hole spin, recombination is slowed down due to angular momentum conservation. We can estimate the effective decoherence rate due to recombination by considering the nuclear spin flip rate. The probability for a single electron to flip-flop with one nucleus is of order $(A^e/\gamma)^2$, so that the nuclear spin flip rate is $(A^e)^2/\gamma\sim0.1$ MHz. We can then describe the electron-hole-nuclear dynamics with a Liouville equation, $\dot\rho=-i[H_{\text{hyp}},\rho]+L[\rho]$, with initial condition
\begin{align}
\rho_{0\mathbf{k}}&=\frac{n^{\infty}_{e\mathbf{k}}}{8}(\sigma_0(s_0+s_z)(\tau_0+\tau_z))\nonumber \\ + &\frac{n^{\infty}_{h\mathbf{k}}}{8}(\sigma_0(s_0-s_z)(\tau_0-\tau_z)),
\end{align}
where $n^{\infty}_{e\mathbf{k}}$ and $n^{\infty}_{h\mathbf{k}}$ are the steady state solutions of the SBEs. The Lindblad term $L[\rho]$ incorporates the effective decoherence effect described above. Choosing this term to be on the order of 1 GHz reproduces the correct 0.1 MHz nuclear spin flip rate. (Additional decoherence and relaxation mechanisms can be incorporated by shifting this value and/or considering additional Kraus operators.) Note that we do not include nuclear spin relaxation since this is on the order of 1 sec \cite{WNMR}. The Liouville-von Neumann equation can be solved exactly to yield an explicit analytical (albeit cumbersome) formula for $\rho(t)$, from which the evolution of the nuclear spin-up probability can be deduced. We find that ${\rho}_\uparrow=f_\uparrow(t) +\rho_\uparrow^\infty$, where $f_\uparrow(t)$ contains the  complicated time dynamics, and $\rho_\uparrow^\infty$ is the steady state value \cite{Havel:2003}(see Appendix C). These solutions then directly give us $r_-$ and, $r_+$,   the nuclear spin-flip rates from spin-up to down, and vice-versa (see Appendix C).

\begin{figure}
\includegraphics[scale=0.190]{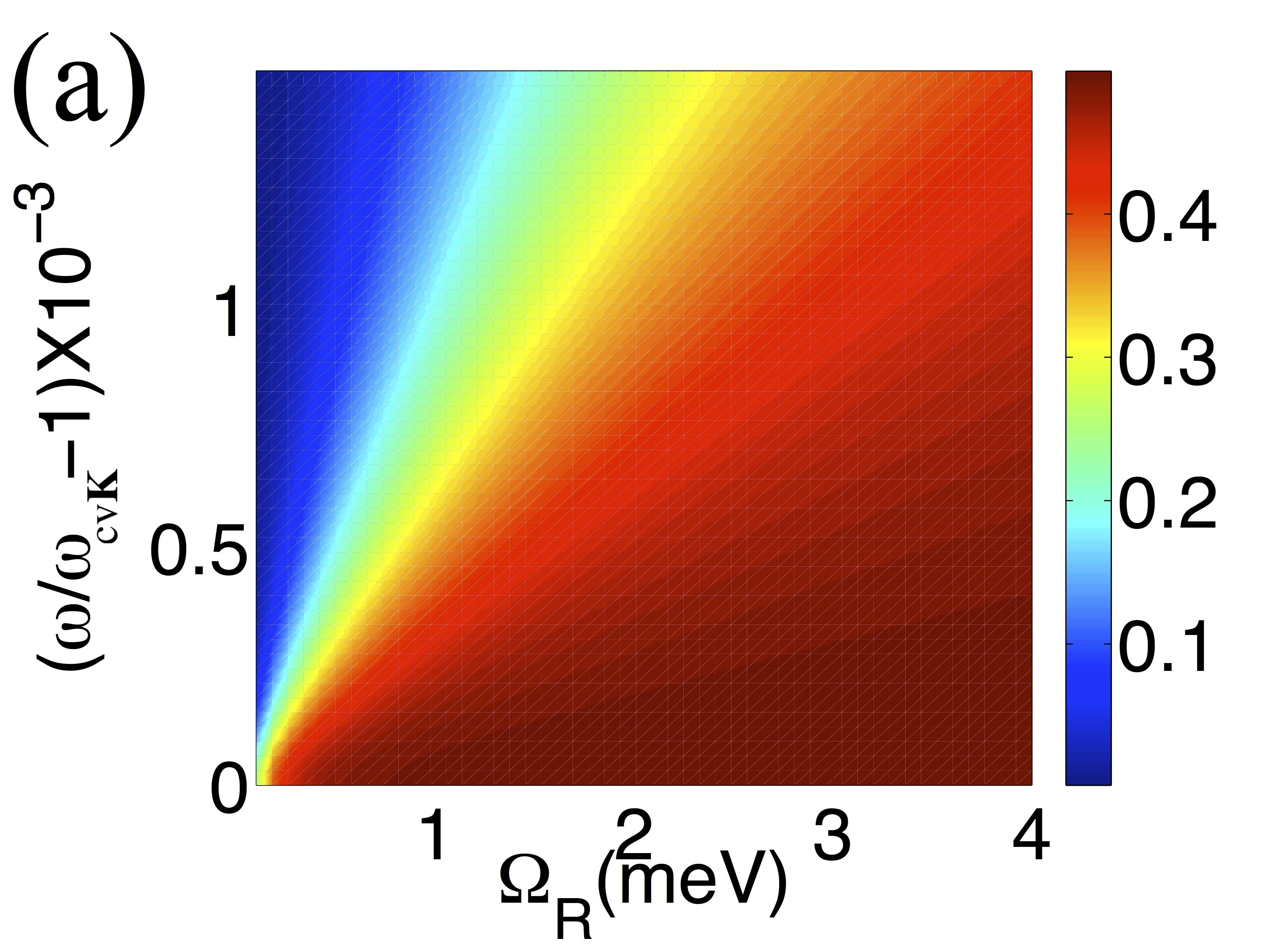}
\includegraphics[scale=0.190]{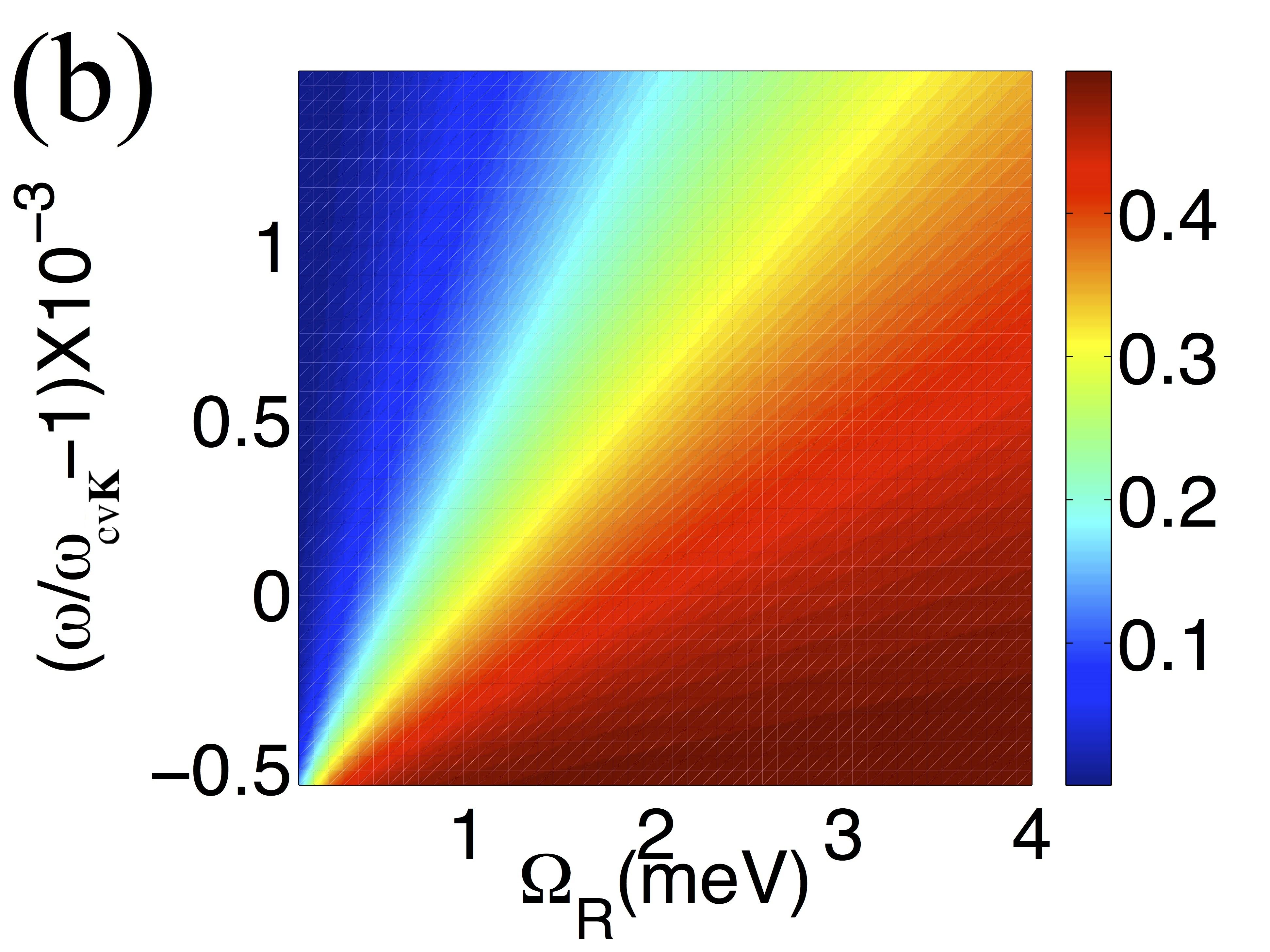}
\caption{(Color online) Density plots for electron distribution function $n^{e\infty}_{\mathbf{K}}$ as a function of laser detuning and Rabi frequency, for a constant $\sigma^+$ drive. \textit{(a)} The case of zero nuclear polarization, $\langle m\rangle=0$. \textit{(b)} The case when steady state polarization has been achieved, with mean nuclear polarization  $\langle m\rangle=0.20 N$, for a sample of $N=10^4$ nuclei. The origin (which is the intersection of red, yellow and blue lines) is shifted from zero on the $y-$axis to $y=-0.5$, since DNP behaves like a detuning parameter.  The hyperfine strengths were chosen to be $A^e=-0.5\mu eV$, $A^h=-1.52\mu eV$, $\eta=0.23$~\cite{WYao:2016}. }
\label{Fig_3}
\end{figure}

Multi-nuclear effects can now be incorporated in the above procedure by including the effective (Overhauser) magnetic field generated by a large number $N$ of nuclei with net polarization $m$. More precisely, we need to calculate the distribution function $P(m)$, which is the probability that the net nuclear spin polarization is $m$. We do this by including the effective Zeeman term, $H_Z=A^ems_z(\tau_0+\tau_z)/2+A^hms_z(\tau_0-\tau_z)/2$, due to the Overhauser field corresponding to arbitrary polarization $m$ in the Liouville-von Neumann equation, and Zeeman shifting the band energies (by $A^{e/h}m/2$) in the steady state distribution functions for $n^{\infty}_{e/h\mathbf{k}}$ (see Appendix C). Solving the Liouville-von Neumann equation (now for $H_{hyp}+H_Z$) now yields $m$-dependent flip rates $r_\pm(m)$, which can be fed into a kinetic equation for the polarization distribution $P(m)$ with the following iterative steady state solution~\cite{Greilich:2007,Barnes:2011, Economou:2014}:
\begin{align}
\frac{P(m)}{P(m-2)}=\frac{N-m+2}{N+m}\frac{r_+(m-2)}{r_-(m)}.
\label{Eq_for_Pm}
\end{align}
The mean value $\langle m\rangle=\sum_m P(m)m$ gives the net DNP generated from optical driving. The steady-state carrier distribution functions obtained from the SBEs are then adjusted to account for the nuclear feedback:
\begin{eqnarray}
n^{\infty}_{e\mathbf{k}}(\delta_\mathbf{k})\rightarrow \sum\limits_m P(m) n^{\infty}_{e\mathbf{k}}(\delta_\mathbf{k}+ m A^e- mA^h).
\label{Eq_elec_feedback_1}
\end{eqnarray}
where $\delta_\mathbf{k}$ is the detuning. In Fig.~\ref{Fig_3} we show a density plot for the electron distribution function $n^{\infty}_{e\mathbf{K}}$ as a function of detuning and Rabi frequency for a sample of $N=10^4$ nuclei illuminated by a constant $\sigma^+$ optical drive. We also show the plot when nuclear polarization is zero ($\langle m\rangle=0$), contrasting it to the case when the nuclear spins are polarized in the steady state, reaching $\langle m\rangle\sim 0.2 N$. We note that when nuclear spins are polarized, the origin of the plot shifts downwards on the detuning axis, clearly demonstrating that DNP acts like an additional detuning parameter.

\section{Experimental proposal} 
\begin{figure}
\includegraphics[scale=0.22]{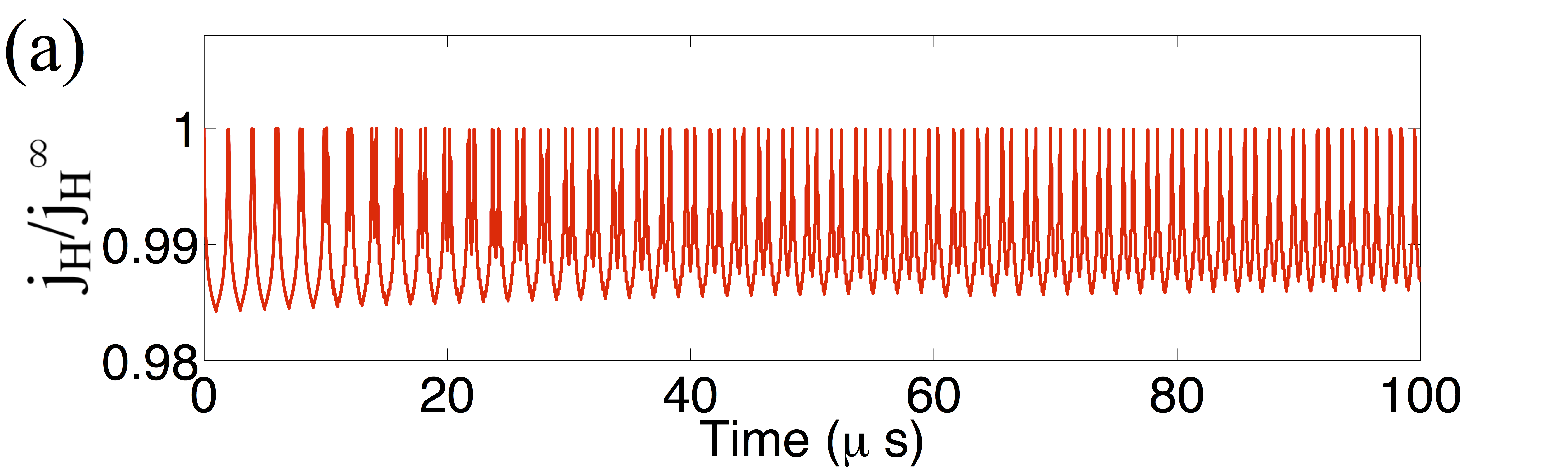}
\includegraphics[scale=0.22]{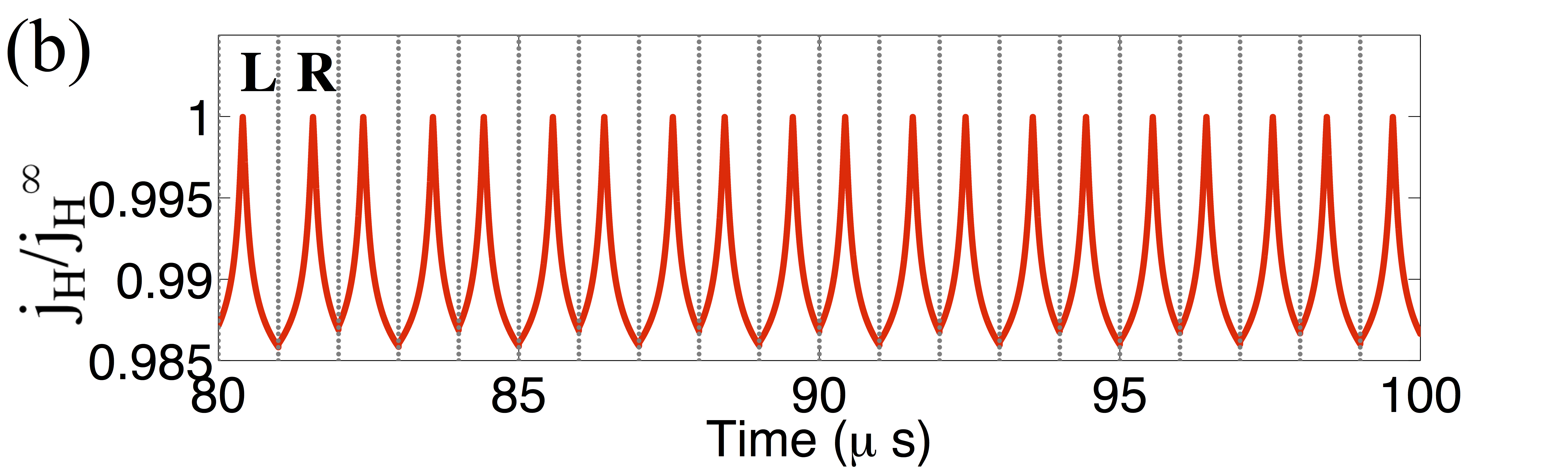}
\includegraphics[scale=0.22]{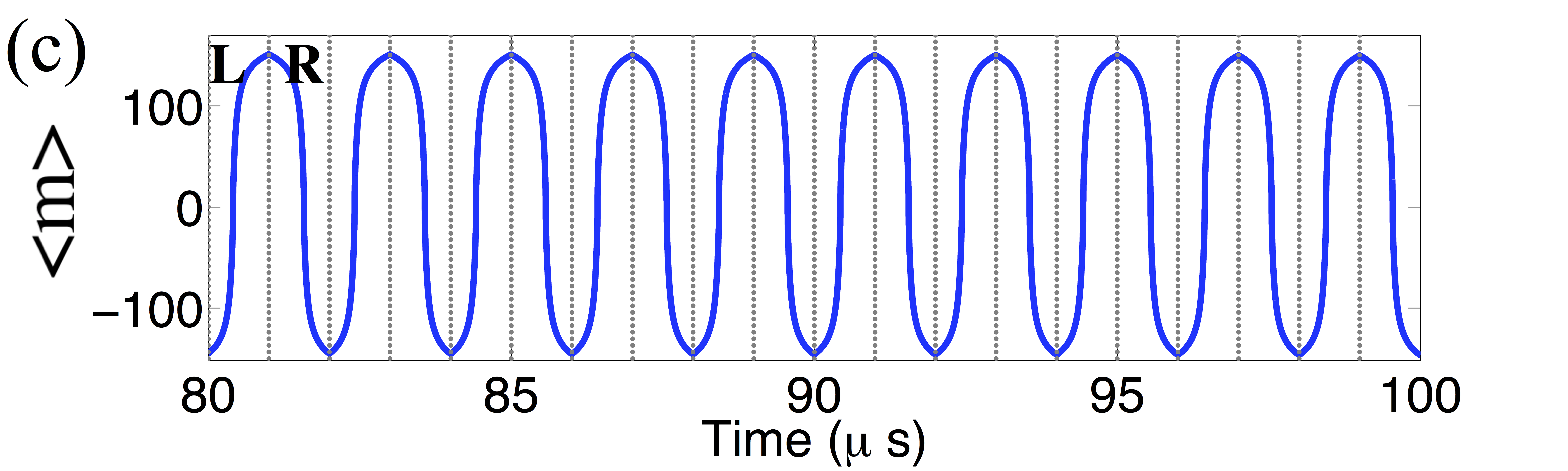}
\includegraphics[scale=0.22]{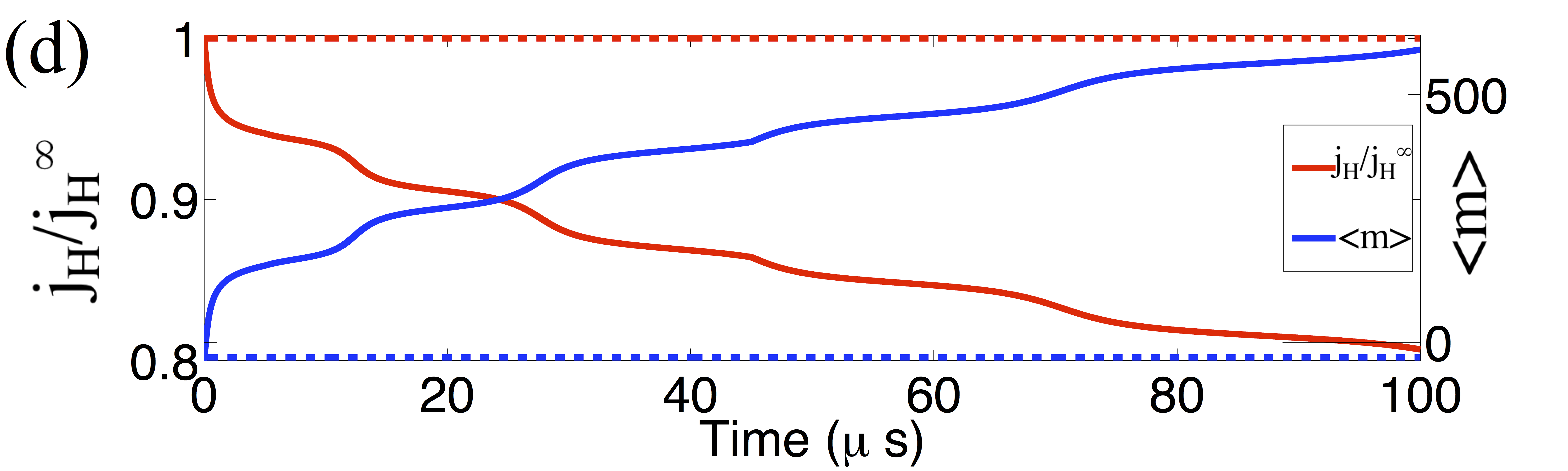}
\caption{(Color online) \textit{(a)} The magnitude of the Hall current under modulated light polarization switching every $1 \mu$s. \textit{(b)} The Hall signal modulation zoomed between $80\mu s-100\mu s$. The dotted lines indicate the polarization switching between LCP and RCP. The signal reaches a steady state and is synchronized with the frequency of polarization switching. \textit{(c)} Plot of dynamical nuclear polarization $m$ zoomed between $80\mu s-100\mu s$. \textit{(d)} Hall current and nuclear magnetization for a constant $\sigma^+$ drive. The dotted red and blue lines on the top and bottom indicate the plots in the absence of DNP. These plots are for the case of $N=10^4$ nuclei, and $\Omega_{R\mathbf{K}}=0.5meV$. The hyperfine values were chosen to be the same as that in Fig.~\ref{Fig_3}. }
\label{Fig_4}
\end{figure}

DNP suppresses the electron band population, and therefore this effect should be detectable in optical and transport measurements. Here we focus on a specific transport response, namely the anomalous Hall response.
As the band population drops, we expect the Hall current to exhibit a similar drop in magnitude (see Fig~\ref{Fig_4}(d)).
We also find that DNP can give rise to a striking modulation in the magnitude of the Hall current when continuous-wave optical driving with an alternating sequence of $\sigma^+$ and $\sigma^-$ polarization is applied, as shown in Fig.~\ref{Fig_4}. We solve for the time evolution of the electron-hole-nuclear system as $\tilde{\rho}(t+dt)=\mathcal{S}(dt)\tilde{\rho}(t)$, where $\mathcal{S}$ (which is in general a function of $n^{\infty}_{e\mathbf{k}}$ and $m$, along with other parameters) is defined in (see Appendix C). Performing a partial trace at every time instant gives $\rho_\uparrow(t)$ and $\rho_\downarrow(t)$, which gives us a measure of nuclear magnetization $\langle m(t)\rangle= N(\rho_\uparrow(t) - \rho_\downarrow(t))$. This strategy then allows us to calculate the time evolution for a system of $N$ nuclei interacting with polarized electron-hole carriers. Since this timescale is much slower than the timescale associated with electron transport, we also calculate $n^{\infty}_{e\mathbf{k}}(t)$ including nuclear feedback at every time step using Eq.~\ref{Eq_elec_feedback_1}, which gives us  $j_{\text{Hall}}$ from Eq.~\ref{Eq_J_HAll}.

In Fig.~\ref{Fig_4} we plot the magnitude of the Hall current for light with periodically modulated polarization. The Hall signal modulation reaches a steady state and is synchronized with the frequency of polarization switching. Note that in the steady state, the signal reaches a maximum and then decreases in intervals of fixed light polarization. This is because the nuclei are first depolarized in the first half of the interval and then polarized in the opposite direction in the second half of the interval. When $j_H=j_H^\infty$, the nuclei are unpolarized ($\langle m\rangle=0$). This particular modulation can be interpreted as a clear signature of the underlying nuclear dynamics. We also contrast this time evolution to the case of a constant $\sigma^+$ drive (i.e. no switching of polarization), also plotted in Fig.~\ref{Fig_4}. The striking difference between the two scenarios is evident from our plots, where in the former case, both DNP and the Hall current saturate to a steady state value. The saturation values are controlled by hyperfine strengths, Rabi frequency, and laser detuning. Our results are for an ensemble of $N=10^4$ nuclei, but one expects a larger signal in the actual experimental scenario where $N\sim 10^6-10^8$, on a timescale relevant to the collective nuclear dynamics. In Appendix E, we confirm that these effects are not modified by the presence of a small conduction band spin-splitting.
\section{Conclusion}
In conclusion, we have shown that DNP has important consequences for optically driven valley polarization. DNP primarily acts like a detuning parameter and alters the band populations via a feedback mechanism. We find a striking modulation of the anomalous Hall current caused by DNP that can be detected via transport measurements in TMDs and can be interpreted as a clear signature of the underlying nuclear bath. Our results serve as a guide for DNP-based experiments in TMDs, uncovering the rich interplay of valley polarization and nuclear spin dynamics in these materials. 

\appendix

\section{Self-consistent Boltzmann solution to carrier dynamics in TMDs}
In order to analyze the dynamics of optically induced carriers in TMDs, we can resort to the solution of a time-dependent Boltzmann transport equation. We will assume that the electron distribution function near a valley point is described by the function $f_{s,\mathbf{k},\mathbf{r}}$, where $s$ is the band index ($s=\pm 1$ for valence and conduction band respectively). The quasiclassical Boltzmann dynamics can be captured by the equation 
\begin{align}
\frac{\partial f_{s,\mathbf{k},\mathbf{r}}}{\partial t} + \dot{\mathbf{k} } \cdot\frac{\partial f_{s,\mathbf{k},\mathbf{r}}}{\partial \mathbf{k}} + \dot{\mathbf{r} } \cdot\frac{\partial f_{s,\mathbf{k},\mathbf{r}}}{\partial \mathbf{r}} = \mathcal{I}(f_{s,\mathbf{k},\mathbf{r}}) 
\label{Eq_Supp_Boltz_0}
\end{align}
In the above equation, $\mathcal{I}(f_{s,\mathbf{k},\mathbf{r}})$ is the collision integral which introduces a relaxation time to the carriers. To describe optically induced carriers, we also need to account for a term which describes dynamical carrier production.  We will first deduce this rate. We will consider the solution to the two level problem where the total Hamiltonian for the system is $H=H_{0,\mathbf{k}}+H_I$, where $H_{0\mathbf{k}}$ is the free fermionic Hamiltonian and $H_I$ is the interaction with electromagnetic field described in the main text. The wavefunction of the Hamiltonian can be expressed as 
\begin{align}
|\psi(t)\rangle = c_v (t)|\psi_v(t)\rangle + c_c(t)|\psi_c(t)\rangle, 
\label{Eq_Supp_psi_t}
\end{align}
where $|\psi_v(t)\rangle = |\psi_v\rangle e^{-iE_v t/\hbar}$, and $|\psi_c(t)\rangle = |\psi_c\rangle e^{-iE_c t/\hbar}$, and $E_{v,c} = \mp E_{\mathbf{k}}$ are the energy bands. Substituting for Eq.~\ref{Eq_Supp_psi_t} in the Schrodinger's equation, we have
\begin{align}
i\hbar \frac{d}{dt} c_v(t) = c_c(t) E(t)d_{\mathbf{k}} e^{-i\omega_{cv\mathbf{k}}t}\\
i\hbar \frac{d}{dt} c_c(t) = c_v(t) E(t)d^*_{\mathbf{k}} e^{+i\omega_{cv\mathbf{k}}t}
\end{align}
where $\omega_{cv\mathbf{k}}$ is the bandgap. For electromagnetic field $E(t) = (E/2)(e^{-i\omega t} + e^{+i\omega t})$, and applying the rotating wave approximation, the above equations transform to 
\begin{align}
i \frac{d}{dt} c_v(t) =\frac{1}{2}\Omega_{R}(\mathbf{k})e^{-i(\omega-\omega_{cv\mathbf{k}})t}c_c(t)\\
i \frac{d}{dt} c_c(t) =\frac{1}{2}\Omega^*_{R}(\mathbf{k})e^{+i(\omega-\omega_{cv\mathbf{k}})t}c_v(t)
\end{align}
where $\Omega_{R}(\mathbf{k})=Ed_{\mathbf{k}}/\hbar$ is the Rabi frequency. When $\omega=\omega_{cv\mathbf{k}}$, the above equations give the usual Rabi oscillations. The rate at which electrons are introduced is given by $d\rho_{cc}/dt$, where $\rho_{cc}=|c_c(t)|^2$, and is given by the following Bloch equations
\begin{align}
\frac{d}{dt}\rho_{cc} &= \frac{-i}{2}\Omega^*_R e^{i(\omega-\omega_{cv\mathbf{k}})t} \rho_{vc} + \frac{i}{2}\Omega_R e^{-i(\omega-\omega_{cv\mathbf{k}})t} \rho_{cv} \nonumber \\
&= -\frac{d}{dt}\rho_{vv}\\
&\frac{d}{dt} \rho_{vc} = \frac{d}{dt} \rho^*_{cv} = \frac{i}{2}\Omega_R e^{-i(\omega-\omega_{cv\mathbf{k}})t} (\rho_{vv}-\rho_{cc})
\label{Eq_Supp_bloch_1}
\end{align}
where $\rho_{vv}=|c_v(t)|^2$, $\rho_{cv}=c^*_v(t)c_c(t)$, $\rho_{vc}=c^*_c(t)c_v(t)$. Further, note that $\rho_{cc}+\rho_{vv}=1$. The above equations have the following general solution for the case when $\rho_{vv}(0)=1, \rho_{cc}(0)=0, \rho_{cv}(0)=0$.
\begin{align}
&\rho_{cc}(t) = \frac{|\Omega_R|^2}{\Omega^2}\sin^2(\Omega t/2)\\
&\rho_{cv}(t) = e^{i(\omega-\omega_{cv\mathbf{k}})t}\frac{\Omega_R}{\Omega^2}\sin(\Omega t/2) (-(\omega-\omega_0)\sin(\Omega t/2)\nonumber \\
&-i\Omega\cos(\Omega t/2)),
\end{align}
where $\Omega = \sqrt{|\Omega_R|^2 + (\omega-\omega_{cv\mathbf{k}})^2}$. In that case the rate at which electrons are introduced is then given by 
\begin{align}
W(\mathbf{k},\tau) = \frac{d\rho_{cc}}{dt} = \frac{|\Omega_R|^2}{2\Omega}\sin(\Omega t)
\end{align}
We now introduce a damping factor $\Gamma$ related to the spontaneous emission. The Bloch equations then become
\begin{align}
&\frac{d}{dt}\rho_{cc} = \frac{-i}{2}\Omega^*_R e^{i(\omega-\omega_{cv\mathbf{k}})t} \rho_{vc} + \frac{i}{2}\Omega_R e^{-i(\omega-\omega_{cv\mathbf{k}})t} \rho_{cv} - 2\Gamma\rho_{cc}\\
&\frac{d}{dt} \rho_{vc} = \frac{d}{dt} \rho^*_{cv} = \frac{i}{2}\Omega_R e^{-i(\omega-\omega_{cv\mathbf{k}})t} (\rho_{vv}-\rho_{cc})-{\Gamma}\rho_{vc}
\label{Eq_Supp_bloch_2}
\end{align}
When the detuning $\delta = \omega-\omega_{cv\mathbf{k}}\neq 0$ and $\Gamma\neq 0$, then it is useful to rewrite the equations in a slightly different form
\begin{align}
&\frac{d}{dt}{\rho_{cc} }= -\frac{i}{2}\Omega_R^* \tilde{\rho}_{vc} + \frac{i}{2}\Omega_R\tilde{\rho}_{cv} - 2\Gamma \rho_{cc}\\
&\frac{d}{dt}\tilde{\rho}_{vc}=\frac{i}{2}\Omega_R(1-2\rho_{cc}) + i\delta \tilde{\rho}_{vc} - \Gamma \tilde{\rho}_{vc}
\end{align}
Denoting $w=1-2\rho_{cc}$, $\tilde{\rho}_{vc} = u+iv$, and $\Omega_R=i\Omega_r$, the equations reduce to the following coupled differential equations 
\begin{align}
&\frac{dw}{dt} = 2\Omega_r u + 2\Gamma - 2\Gamma w\\
&\frac{du}{dt} = -\frac{\Omega_r w}{2} - \delta v - \Gamma u\\
&\frac{dv}{dt} = -\Gamma v + \delta u
\end{align}
We Laplace transform the above equations, along with the initial conditions $(w(0),u(0),v(0))=(+1,0,0)$. 
\begin{align}
& sZ - 1 - 2\Omega_r X - \frac{2\Gamma}{s} + 2\Gamma Z = 0\\
& sX + \frac{\Omega_r}{2} Z + \delta Y + \Gamma X=0\\
& sY + \Gamma Y - \delta X = 0
\end{align}
where $X$, $Y$, and $Z$ are the Laplace transforms of $u$, $v$ and $w$ respectively. The above set of equations can be solved for
\begin{align}
Z = \frac{((2\Gamma + s)(\delta^2 + \Gamma^2 + 2\Gamma s + s^2))}{s\Delta(s)}
\end{align}
where $\Delta(s) = 2\Gamma^3 + 5\Gamma^2 s+\Gamma\Omega^2 + 2\Gamma\delta^2 + 4\Gamma s^2 + \Omega^2 s + \delta^2 s + s^3$. Similarly one may also solve for $X$ and $Y$. The equation $\Delta(s)=0$ has three roots of the form $s=-a$, $s=-m\pm in$. We can now solve for $w(t)$ using the inverse Laplace transform, which gives us the following expression
\begin{align}
w(t) = Ae^{-at} + Be^{-mt} \cos(nt) + (C/n) e^{-mt} \sin(nt) + D
\label{Eq_Supp_w_t_1}
\end{align} 
where \begin{align}
A = -\frac{(2\Gamma- a)(\delta^2 + \Gamma^2 - 2\Gamma a + a^2)}{a((m-a)^2 +n^2)} 
\end{align}
\begin{align}
D = \frac{2\Gamma (\delta^2 + \Gamma^2)}{a(m^2 + n^2)}
\end{align}
\begin{align}
B = 1-A-D; C = Aa + Bm
\end{align}
\begin{align}
-a = -\frac{4\Gamma}{3} - \frac{2^{1/3} x}{3(y + \sqrt{4x^3+y^2})^{1/3}} + \frac{(y+\sqrt{4x^3 + y^2})^{1/3}}{3\times 2^{1/3}}
\end{align}
\begin{align}
-m+in&=-\frac{4\Gamma}{3} + \frac{(1+\sqrt{3}i)x}{3\times 2^{2/3}(y+\sqrt{4x^3 + y^2})^{1/3}} \nonumber \\
- &\frac{(1-\sqrt{3}i)(y+\sqrt{4x^3 + y^2})^{1/3}}{6\times 2^{1/3}}
\end{align}
where $x=3\delta^2 - \Gamma^2 + 3\Omega_R^2$, and $ y = -18 \delta^2 \Gamma - 2\Gamma^3 + 9\Gamma\Omega_R^2$.

The final rate is then given by 
\begin{align}
&W(\mathbf{k},\tau,t) = -\frac{1}{2}\frac{dw}{dt} = \frac{1}{2}(-Aa e^{-at} - Bme^{-mt} \cos(nt) \nonumber\\
&- Bn e^{-mt}\sin(nt) - (Cm/n) e^{-mt} \sin(nt) - Ce^{-mt}\cos(nt))\nonumber\\
\label{Eq_Supp_W_exciton_3}
\end{align}
We will now use this result for the solution of Boltzmann equation. Assuming spatially uniform (along the plane of the sample) fields, the Boltzmann equation (~\ref{Eq_Supp_Boltz_0}) becomes
\begin{align}
\frac{\partial f_{s,\mathbf{k}}}{\partial t} + \dot{\mathbf{k} } \cdot\frac{\partial f_{s,\mathbf{k}}}{\partial \mathbf{k}} = \mathcal{I}(f_{s,\mathbf{k}}) + U^s(\mathbf{k},\tau,t) (f_{s',\mathbf{k}}-f_{s,\mathbf{k}})
\label{Eq_Supp_Boltz_1}
\end{align}
where $\mathcal{I}(f_{s,\mathbf{k}})$ is the collision integral which describes the effect of carrier scattering. $U(\mathbf{k},\tau,t)$ is related to the rate at which carriers are introduced discussed previously. 
\begin{align}
U^s(\mathbf{k},\tau,t) = \frac{W(\mathbf{k},t,\tau)}{w^s(t)},
\label{Eq_Supp_U_W_rel_1}
\end{align}
where $W(\mathbf{k},\tau,t)$ is given in Eq.~\ref{Eq_Supp_W_exciton_3} and $w^c(t)=1-2\rho_{cc}(t)$ in Eq.~\ref{Eq_Supp_w_t_1} for conduction band, and $w^v(t)=1-2\rho_{vv}(t)$ for valence band.
\begin{widetext}
\begin{align}
U^s(\mathbf{k},\tau,t) = \frac{\frac{1}{2}(-Aa e^{-at} - Bme^{-mt} \cos(nt) - Bn e^{-mt}\sin(nt) - (Cm/n) e^{-mt} \sin(nt) - Ce^{-mt}\cos(nt))}{Ae^{-at} + Be^{-mt} \cos(nt) + (C/n) e^{-mt} \sin(nt) + D}
\end{align}
\end{widetext}
In the absence of carrier scattering, and zero static electric and magnetic fields, the Boltzmann equation simplifies to 
\begin{align}
\frac{\partial f_{s,\mathbf{k}}}{\partial t} =  U^s(\mathbf{k},\tau,t) (f_{s',\mathbf{k}}-f_{s,\mathbf{k}})
\end{align}
For the conduction band, the above equation is 
\begin{align}
\frac{\partial f_{c,\mathbf{k}}}{\partial t} =  U^c(\mathbf{k},\tau,t) (1-2f_{c,\mathbf{k}}) = W(\mathbf{k},\tau,t)
\end{align}
as expected from the solution to the Bloch equations. 

\begin{figure}
\includegraphics[scale=0.18]{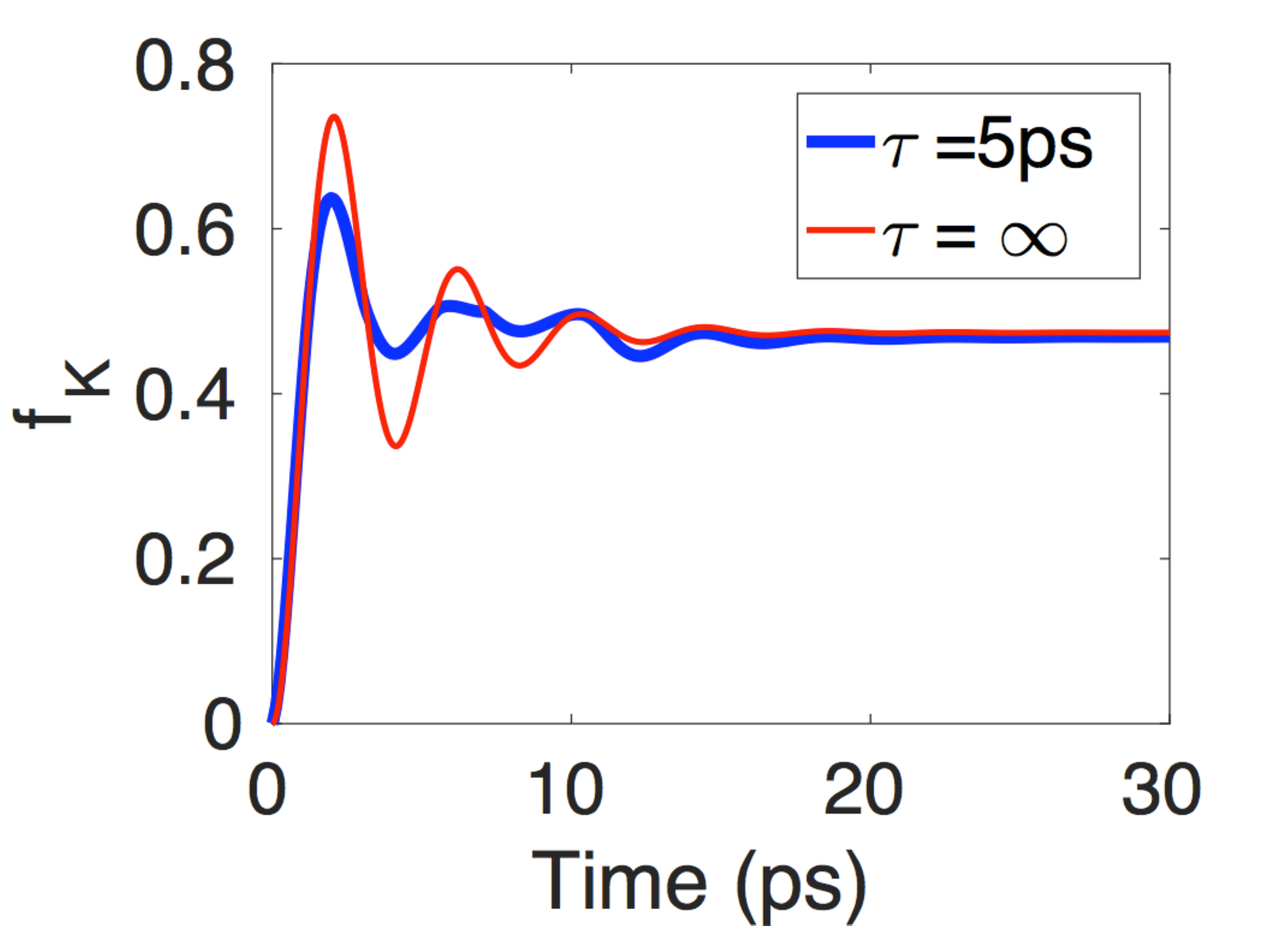}
\includegraphics[scale=0.18]{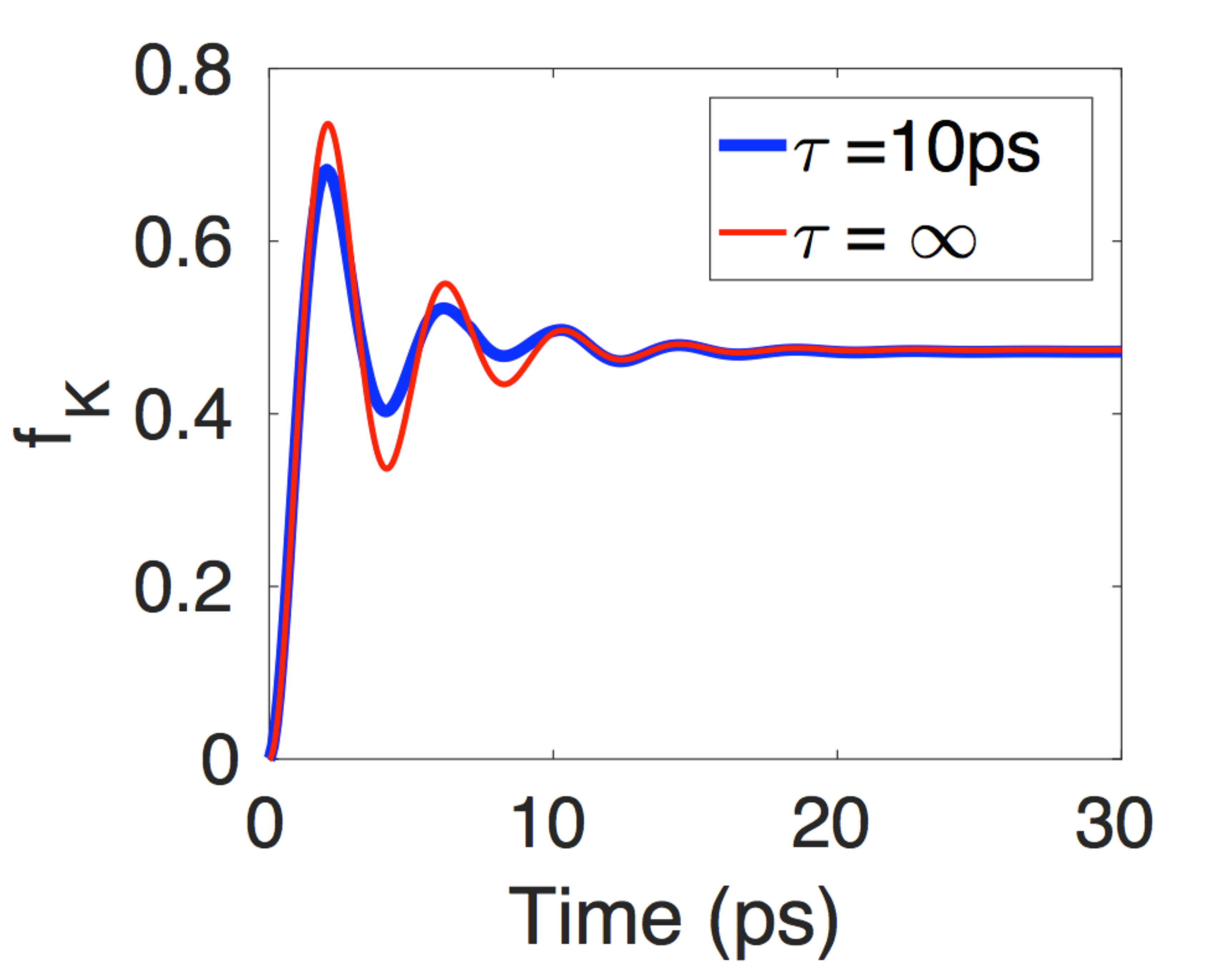}
\includegraphics[scale=0.18]{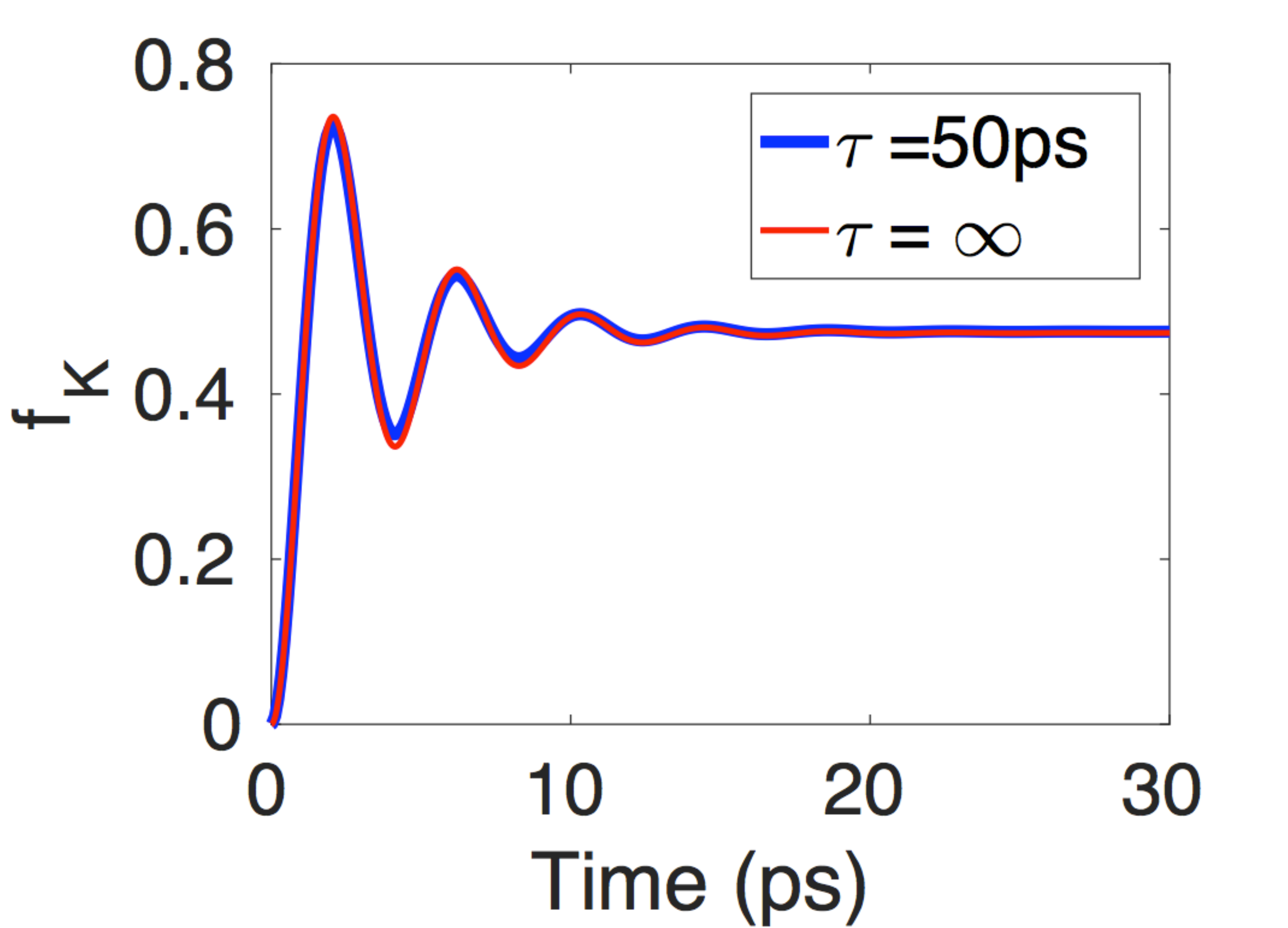}
\caption{The electron distribution function $f_{\mathbf{K}}$ at a valley point as a function of time obtained from the Boltzmann equation, for various values of scattering times ($\tau$). In red is the distribution function plotted in the absence of any scattering ($\tau\rightarrow\infty$). For large $\tau$ values, the electron distribution function approaches the red curve. We use $1/\Gamma = 5ps$, $\omega_{R\mathbf{K}}=1meV$, $\delta=0.1\omega_{R\mathbf{K}}$.}
\label{Fig_Supp_f_K_vs_time_various_taus}
\end{figure}

Using a phenomenological description for carrier relaxation we have 
\begin{align}
\mathcal{I}(f_{s,\mathbf{k}}) = -\frac{f_{s,\mathbf{k}}-f^0_{s,\mathbf{k}}}{\tau_{s,\mathbf{k}}}
\end{align}
where $f^0_{s,\mathbf{k}}$ is the equilibrium distribution function, and $\tau_{s,\mathbf{k}}$ is the phenomenological intra-band scattering time. Note that now the equilibrium distribution function $f^0_{s,\mathbf{k}}$ must be band dependent. When $t\rightarrow \infty$, $U(\mathbf{k},\tau,t)\rightarrow 0$, as a steady state has been reached, with hole and electron type carriers in the valence band and conduction band respectively. The function $f^0_{s,\mathbf{k}}$ can be identified with the electron distribution in the $s$ band for a finite carrier distribution created by the laser i.e. $f^0_{c,\mathbf{k}}\equiv \rho_{cc}(t\rightarrow\infty)$ and $f^0_{v,\mathbf{k}}\equiv \rho_{vv}(t\rightarrow\infty)$, which holds exactly true in the limit $\tau\rightarrow\infty$. For a generic $\tau$ value this might not hold true always. Therefore, for numerical calculation of $f_{\mathbf{k}}$, we start with an initial guess of $f^0$, and then compare $f_{\mathbf{k}}(t\rightarrow\infty)$ with the guess $f^0$, and re-evaluate $f_{\mathbf{k}}$ until the condition $f_{\mathbf{k}}(t\rightarrow\infty)\rightarrow f^0$ is satisfied. 

The following equation describes the distribution function in the conduction band 
\begin{align}
&\frac{\partial f_{c,\mathbf{k}}}{\partial t} + \dot{\mathbf{k} } \cdot\frac{\partial f_{c,\mathbf{k}}}{\partial \mathbf{k}} = -\frac{f_{c,\mathbf{k}}-f^0_{c,\mathbf{k}}}{\tau_{c,\mathbf{k}}} + U(\mathbf{k},\tau,t) (1-2f_{c,\mathbf{k}}) 
\end{align}
We first discuss solutions of the above equation in the absence of any static electric fields. i.e. $\dot{\mathbf{k}}=0$. We now have two rates ($\tau^{-1}_{c,\mathbf{k}}$ and $U(\mathbf{k},\tau,t)$) which govern the dynamics of $f^c_{\mathbf{k}}$, given by
\begin{align}
&\frac{\partial f_{c,\mathbf{k}}}{\partial t}  = -\frac{f_{c,\mathbf{k}}-f^0_{c,\mathbf{k}}}{\tau_{c,\mathbf{k}}} + U(\mathbf{k},\tau,t) (1-2f_{c,\mathbf{k}}) 
\label{Eq_Supp_Boltz_5}
\end{align}

\section{The solution to Semiconductor Bloch Equations (SBEs) in TMD}
Here we will discuss the general solution to SBEs relevant to TMDs. The total Hamiltonian ($H_{\mathbf{k}}+H_I$) described in the main text can be written as 
\begin{align}
H &= \sum\limits_\mathbf{k} {E_{1\mathbf{k}}c^\dagger_{1\mathbf{k}}c_{1\mathbf{k}} + E_{2\mathbf{k}}c^\dagger_{2\mathbf{k}}c_{2\mathbf{k}}} \nonumber \\ 
& +\frac{1}{2}\sum\limits_{\mathbf{k}\mathbf{k}'\mathbf{q}} V_{\mathbf{q}}(c^\dagger_{1\mathbf{k}+\mathbf{q}}c^\dagger_{1\mathbf{k}'-\mathbf{q}}c_{1\mathbf{k}'}c_{1\mathbf{k}} +c^\dagger_{2\mathbf{k}+\mathbf{q}}c^\dagger_{2\mathbf{k}'-\mathbf{q}}c_{2\mathbf{k}'}c_{2\mathbf{k}} \nonumber \\
&+ 2V_{\mathbf{q}}c^\dagger_{1\mathbf{k}+\mathbf{q}}c^\dagger_{2\mathbf{k}'-\mathbf{q}}c_{2\mathbf{k}'}c_{1\mathbf{k}})\nonumber \\
&-\sum\limits_\mathbf{k}\mathcal{E}(t) (d_\mathbf{k} c^\dagger_{1\mathbf{k}}c_{2\mathbf{k}} + h.c.)
\label{Eq_Supp_Full_H_1}
\end{align}  
where the subscript (1,2) is the two band index, $E_{i\mathbf{k}}$ gives the non-interacting band structure as described in the main text, $c^\dagger_{i\mathbf{k}}$ is the electron creation operator, $\mathcal{E}(t)$ is the coupling to electromagnetic field, $d_\mathbf{k}$ is the interband dipole matrix element, and $V(\mathbf{q})$ is the Fourier transform of the Coulomb interaction in two dimensions. 
In order to discuss time-evolution of carriers induced by the laser, we will focus on the following average values: $n_{e\mathbf{k}}=\langle c^\dagger_{c,\mathbf{k}} c_{c,\mathbf{k}}\rangle$, $n_{h\mathbf{k}}=\langle c_{v,\mathbf{k}} c^\dagger_{v,\mathbf{k}}\rangle$, $Q_{\mathbf{k}}=\langle c^\dagger_{v\mathbf{k}}c_{c\mathbf{k}}\rangle$, which denote the band-population in conduction band, valence band, and the inter-band polarization respectively. The coupled Heisenberg equations of motion for the total Hamiltonian (Eq.~\ref{Eq_Supp_Full_H_1}) can be solved where we split the four-operator terms into products of densities and interband polarizations plus the unfactorized rest. Thus we can separate the equation into the Hartree–Fock and scattering parts. The coupled equations reduce to the following as quoted in the main text
\begin{align}
\frac{d Q_{\mathbf{k}}}{dt} &= -i(e_{e\mathbf{k}}+e_{h\mathbf{k}})Q_{\mathbf{k}} - i(n_{e\mathbf{k}}+n_{h\mathbf{k}}-1)\omega_{r\mathbf{k}} + \left[\frac{d Q_{\mathbf{k}}}{dt}\right]_{\text{scatt}} \label{Eq_Supp_Pk_1}\\
\frac{d n_{e\mathbf{k}}}{dt} &= -2 \text{Im} (\omega_{r\mathbf{k}}Q_{\mathbf{k}}^*)+ \left[\frac{d n_{e\mathbf{k}}}{dt}\right]_{\text{scatt}}\label{Eq_Supp_nek_1}\\
\frac{d n_{h\mathbf{k}}}{dt} &= -2 \text{Im} (\omega_{r\mathbf{k}}Q_{\mathbf{k}}^*)+ \left[\frac{d n_{h\mathbf{k}}}{dt}\right]_{\text{scatt}} \label{Eq_Supp_nhk_1}
\end{align}
where $e_{e(h)\mathbf{k}}$ describe the electron(hole) renormalized single-particle energies, and $\omega_{r\mathbf{k}}$ is the generalized Rabi frequency. 
\begin{align}
&e_{i\mathbf{k}} = \epsilon_{i\mathbf{k}} - \sum\limits_{\mathbf{q}}V_{|\mathbf{k}-\mathbf{q}|} n_{i\mathbf{q}} \\
&\epsilon_{e\mathbf{k}} = E_{c\mathbf{k}} \\
&\epsilon_{h\mathbf{k}} = -E_{h\mathbf{k}} +\sum\limits_\mathbf{q} V_{\mathbf{q}} \\
& \omega_{r\mathbf{k}} = \frac{1}{\hbar} \left(d_{\mathbf{k}}\mathcal{E}(t) + \sum\limits_{\mathbf{q}\neq \mathbf{k}}V_{|\mathbf{k}-\mathbf{q}|}Q_{\mathbf{q}}\right)
\end{align}
In Eq.~\ref{Eq_Supp_Pk_1}-\ref{Eq_Supp_nhk_1}, the scattering terms $[d\langle A\rangle/dt]_{\text{scatt}}$ on the right denote the difference between the full derivatives and the Hartree-Fock terms. In principle we can solve the above set of equations numerically. However to make our model analytically tractable, we will assume a phenomenological description of scattering terms, and the limit $V(\mathbf{q})\rightarrow 0$, thus ignoring Coulomb interactions. Since the primary effect of Coulomb interactions is to renormalize single particle energies and the Rabi frequency, this assumption does not change any of our conclusions, at least qualitatively. For the scattering terms, we could also introduce a microscopic theory for phonons and carrier-carrier scattering, but we content ourselves with a phenomenological description which suffices the purpose of this work. We introduce two timescales: $\gamma$ for inter-band depolarization, and $\Gamma$ for intra-band carrier scattering. Since the inter valley scattering rate is much slower than $\Gamma$, we can safely ignore such processes here. For $\Gamma\ll\gamma$, even a self consistent solution to the Boltzmann transport equation can effectively describe band-populations as discussed in the previous section, but the present approach works for the general case.  

Denoting $Q_{\mathbf{k}}=u_\mathbf{k}+iv_\mathbf{k}$, and $w_\mathbf{k}=1-2n_{e\mathbf{k}}$, Eqs.~\ref{Eq_Supp_Pk_1}-\ref{Eq_Supp_nhk_1} can then be rewritten as
\begin{align}
\frac{du_{\mathbf{k}}}{dt} &= \delta_{\mathbf{k}}v_{\mathbf{k}}-\omega_{r\mathbf{k}}w_{\mathbf{k}}-\gamma u_{\mathbf{k}}\\
\frac{dv_{\mathbf{k}}}{dt} &= -\delta_{\mathbf{k}}u_{\mathbf{k}}-\gamma v_{\mathbf{k}}\\
\frac{dw_{\mathbf{k}}}{dt} &= 4\omega_{r\mathbf{k}} - (2\gamma+\Gamma)(1-w_{\mathbf{k}})
\end{align}
where $\delta_{\mathbf{k}} = \omega-\epsilon_{e\mathbf{k}}-\epsilon_{h\mathbf{k}}$ is the laser detuning. The SBEs (Eq.~\ref{Eq_Supp_Pk_1}-\ref{Eq_Supp_nhk_1}) yield the following analytical solution (obtained via the same strategy we used to get the analytical solution of the usual Bloch equations discussed previously)
\begin{align}
w_{\mathbf{k}}(t) &= A_1e^{-at} + A_2e^{-mt} \cos(nt) \nonumber \\
&+ (A_3/n) e^{-mt} \sin(nt) + A_4 
\label{Eq_Supp_w_t_2}
\end{align} 
where
\begin{align}
A_1 &= -\frac{(\Gamma + 2\gamma)(\delta_{\mathbf{k}}^2 + \gamma^2 - 2\gamma a + a^2)}{(a ((m-a)^2 + n^2))}\\
A_4 &= \frac{(\Gamma + 2\gamma)}{(\delta_{\mathbf{k}}^2 + \gamma^2)/(a(m^2+n^2))}\\
A_2 &= 1-A_1-A_4\\
A_3 &= a A_1 + m A_2
\end{align}
and $-a$, $-m\pm n$ are solutions of the equation  $\Delta(s)=0$, where $\Delta(s)$ is defined below
\begin{align}
\Delta(s) &= 2\delta_\mathbf{k}^2 \gamma + \delta_\mathbf{k}^2 s + \Gamma\delta_\mathbf{k}^2 + 2\gamma^3 + 5\gamma^2 s + \Gamma\gamma^2 + 4\gamma\omega_{r\mathbf{k}}^2 \nonumber \\
&+ 4\gamma s^2 + 2\Gamma\gamma s + 4\omega_{r\mathbf{k}}^2 s + s^3 + \Gamma s^2 
\end{align}
\section{Coupling to nuclei-Dynamic Nuclear Polarization}
Since the nuclei simultaneously interact with both electron and holes (with different hyperfine couplings), one needs to consider a density matrix involving electrons, holes and nuclei. For the case of a single nucleus, the hyperfine interaction is given by the Hamiltonian 
\begin{align}
H_{\text{hyp}}&=\frac{A^e}{2} \sigma_z s_z (\tau_0+\tau_z) + \frac{A^h}{2} \sigma_z s_z (\tau_0-\tau_z)\nonumber \\&+ \frac{\eta A^e}{4} (\sigma_x s_x +\sigma_y s_y)(\tau_0+\tau_z) 
\label{Eq_Supp_H_hyp_elechole}
\end{align} 
which written in the basis $(\rho^c_{\uparrow\uparrow},\rho^c_{\uparrow\downarrow},\rho^c_{\downarrow\uparrow},\rho^c_{\downarrow\downarrow}, \rho^v_{\uparrow\uparrow},\rho^v_{\uparrow\downarrow},\rho^v_{\downarrow\uparrow},\rho^v_{\downarrow\downarrow})$, where $c$ and $v$ refer to conduction and valence band respectively.
The initial condition obtained from SBEs is given by
\begin{align}
\rho_{0\mathbf{k}}=\frac{n^{e\infty}_{\mathbf{k}}}{8}(\sigma_0(s_0+s_z)(\tau_0+\tau_z))\frac{n^{h\infty}_{\mathbf{k}}}{8}(\sigma_0(s_0-s_z)(\tau_0-\tau_z))
\label{Eq_Supp_rho_0_elechole}
\end{align}
The solution to the above Hamiltonian is given by the Liouville-von Neumann  equation with the relaxation superoperator term $\dot\rho=-i[H_{\text{hyp}},\rho]+L[\rho]$. We consider the following Lindblad relaxation operator 
\begin{align}
L[\rho] = \sum\limits_{ij} \left(L_{ij}\rho L_{ij}^\dagger-\frac{1}{2}[L_{ij}^\dagger L_{ij}\rho + \rho L_{ij}^\dagger L_{ij}] \right)
\label{Eq_Supp_Lindblad}
\end{align}
The operators $L_{ij}$ are described by $ L_{ij}=\alpha_{ij}|i\rangle\langle j|$.  For effective decoherence effect, the matrix $\alpha=b\mathcal{I}$, is chosen to be a constant times identity operator. The decoherence rate is taken to be around $~\sim 1GHz$, which reproduces the correct rate of nuclear spin-flip. This however does not include the nuclear spin relaxation since $T_{1nuc}$ is on the order of $~\sim 1 sec$, which is much greater than all the other timescales. Additional decoherence effects can also be modeled by shifting this value and/or constructing more Kraus operators.

The Liouville-von Neumann  equation with the relaxation superoperator term can be solved exactly as 
$ \tilde{\rho}(t) = \mathcal{S}(t)\tilde{\rho}(0)$, where $\tilde{\rho}(t)$ is the density matrix $\rho(t)$ written in as a single column vector, and $\mathcal{S}(t)$ is $ \mathcal{S}(t) = e^{(\mathcal{H+G})t}$, where $ \mathcal{H} = -i(H\otimes I - I\otimes H)$, $ \mathcal{G} = \sum_{m}\left[L_m\otimes L_m - \frac{1}{2}I\otimes L_m^\dagger L_m - \frac{1}{2}L_m^\dagger L_m \otimes I\right]$. The analytical solution to $\rho(t)$ is possible to write because we just need the eigenvectors and eigenvalues of $\mathcal{H}+\mathcal{G}$, however the exact form is tedious and hence is not provided here.

\begin{figure}
\includegraphics[scale=0.35]{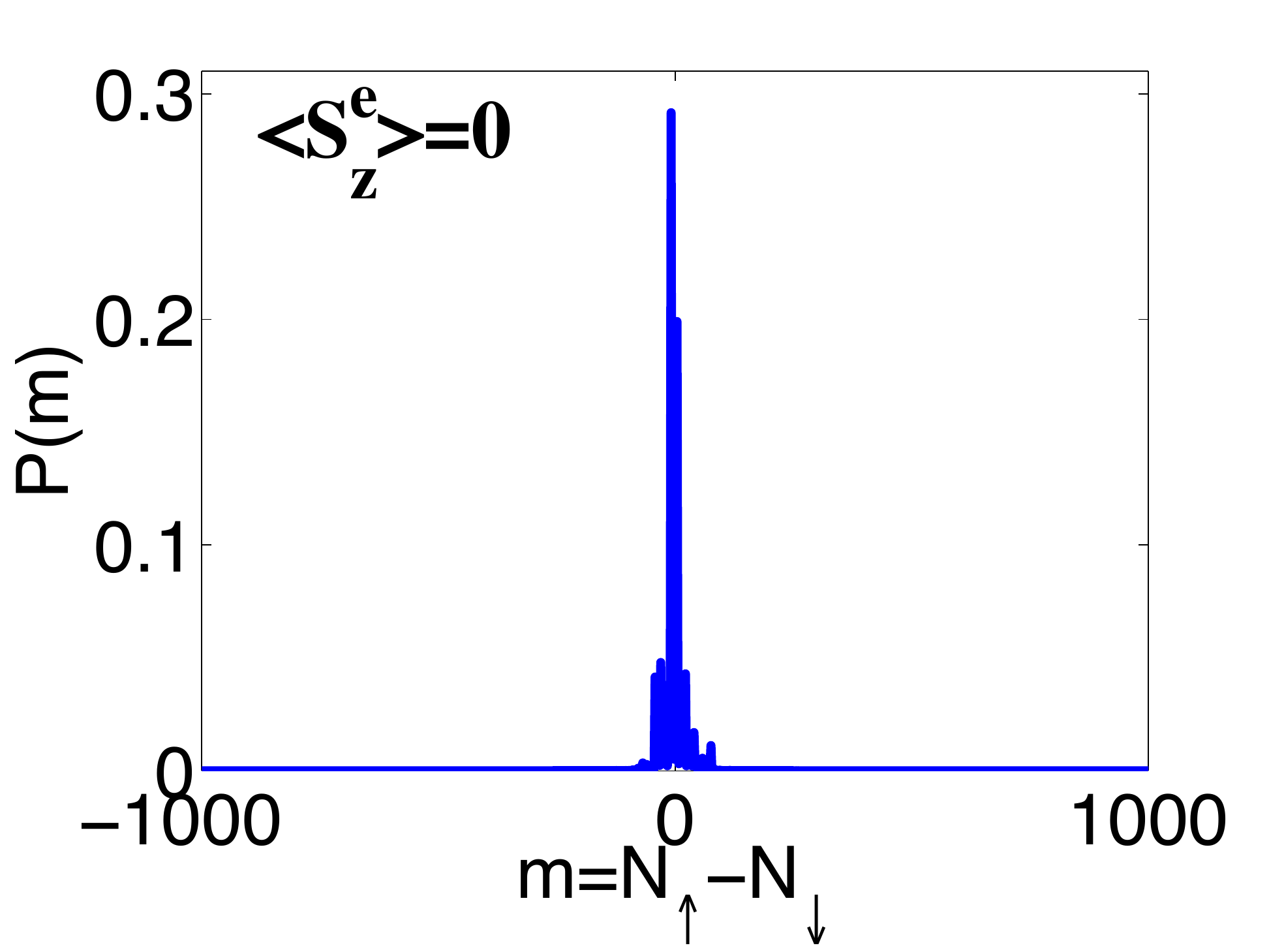}
\includegraphics[scale=0.35]{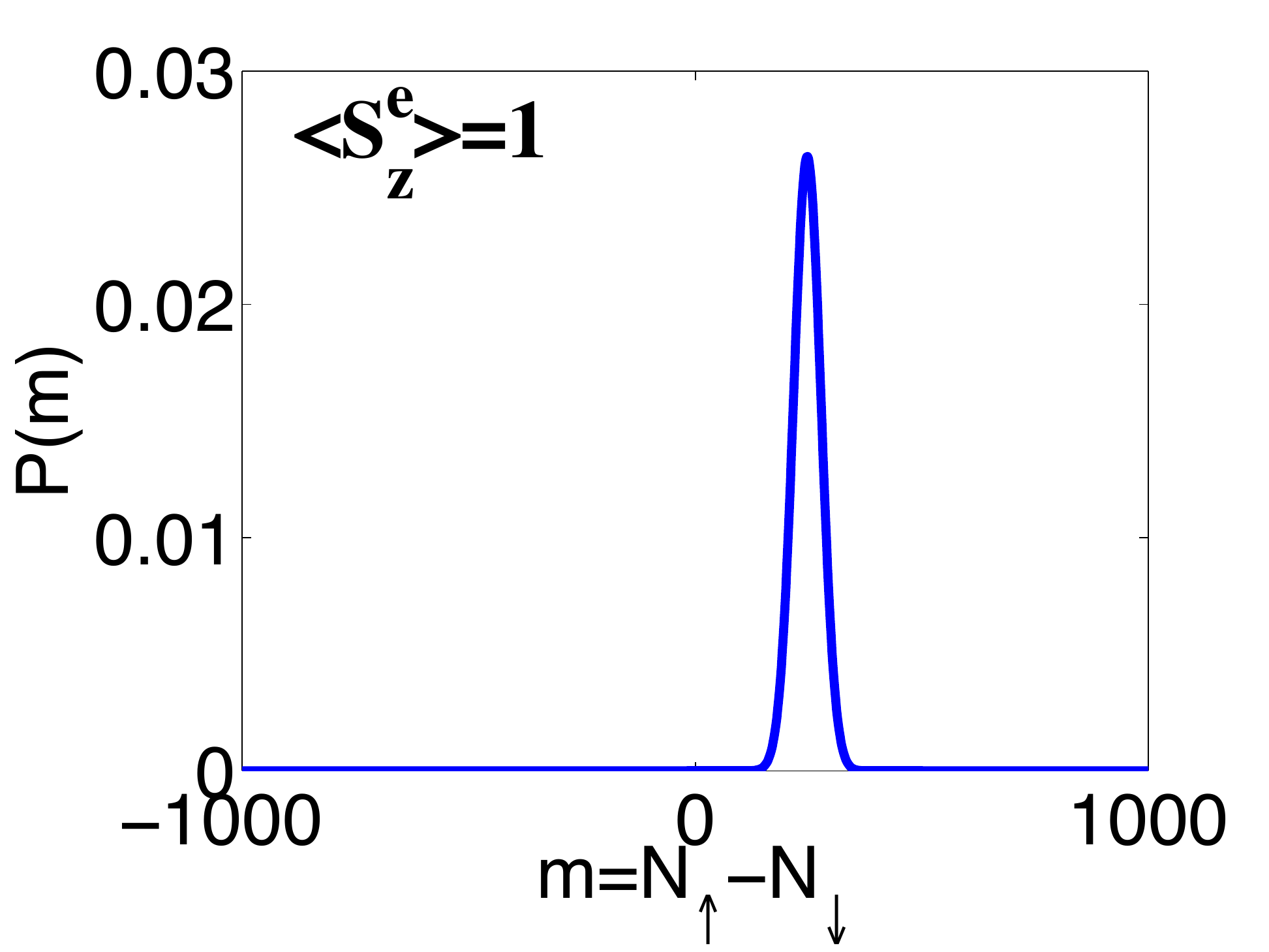}
\caption{Nuclear polarization distribution via DNP for an ensemble of 1000 nuclei. \textit{Top:}  Electrons which have a zero average spin result in a mean nuclear polarization close to zero. \textit{Bottom:} Spin-polarized electrons (as generated in TMDs where we excite a single valley) result in a non-zero nuclear polarization.}
\label{Fig_Supp_2}
\end{figure}

We will now obtain the nuclear flip rates with the help of simplified rate equations. For the case of a single nucleus we may write the following rate equation 
\begin{align}
 \frac{dP_\uparrow}{dt}=-r_-P_\uparrow + r_+ P_\downarrow
 \end{align}
where $P_{\uparrow(\downarrow)}$ is the probability of the nucleus being in the spin-$\uparrow(\downarrow)$ state, and $r_{+/-}$ is the rate of flipping the the nuclear spin from $\uparrow$ to $\downarrow$ or viceversa. Therefore 
\begin{align}
\frac{dP_\uparrow}{dt}=-r_-P_\uparrow + r_+ (1-P_\uparrow)\nonumber\\
\frac{dP_\uparrow}{dt}=-(r_-+r_+)P_\uparrow + r_+ 
\end{align}
This yields the following  solution 
\begin{align}
P_\uparrow(t) &= \left[P_\uparrow(0) - \frac{r_+}{r_-+r_+}\right]e^{-(r_-+r_+)t}+\frac{r_+}{r_-+r_+}\\
 P_\downarrow(t) &= \left[P_\downarrow(0) - \frac{r_-}{r_-+r_+}\right]e^{-(r_-+r_+)t}+\frac{r_-}{r_-+r_+} 
\end{align}
Comparing these solutions with the steady solutions to the Liouville-von Neumann equation, gives us $r_+$ and $r_-$ as $ {r_+}/{r_-+r_+} = \rho^\infty_{\uparrow}$, ${r_-}/{r_-+r_+} = \rho^\infty_{\downarrow}$. Further, the total rate $\lambda=r_++r_-$, is also determined from the analytical solution to the Liouville-von Neumann equation discussed above. Therefore $ r_+ = \lambda \rho_\uparrow^\infty$, $r_- = \lambda \rho_\downarrow^\infty$.
In presence of a multiple nuclei, one needs to calculate the net polarization distribution $P(m)$ for the multi-nuclear system. The constraint on $P(m)$ is that $ \sum\limits_{m=-N}^{m=+N} P(m) = 1$,  where $N$ is the number of nuclei and $m$ represents the difference between spin-up ans spin-down nuclei. $P(m)$ has the following general steady state relation as discussed in the main text
\begin{align}
P(m)=\frac{N-m+2}{N+m}\frac{r_+(m-2)}{r_-(m)}P(m-2)
\label{Eq_Supp_for_Pm}
\end{align} 
which maybe iteratively solved for $P(m)$, once the rates $r_+(m)$ and $r_-(m)$ are obtained in the presence of nuclear polarization. 

In order to calculate $r_\pm(m)$, we first solve the Liouville-von Neumann equation for $\rho(t)$ (described by the above procedure) for the Hamiltonian $ H_Z + H_{hyp} $, 
where $H_Z=A^ems_z(\tau_0+\tau_z)/2+A^hms_z(\tau_0-\tau_z)/2$ incorporates effects due to the Overhauser field in the presence of non-zero $m$. We then obtain $r_\pm(m)$ with the analytical solution to $\rho(t)$ and comparing it to the rate equations (similar to the case of a single nucleus discussed above).  We find that $ r_+(m) = \lambda(m) \rho_\uparrow^\infty$, $ r_-(m) = \lambda(m) \rho_\downarrow^\infty$.
where the total rate $\lambda(m)$ depends on $m=N_\uparrow-N_\downarrow$, and $\rho_{\uparrow\downarrow}^\infty$ depends on the initial conditions and system parameters. Fig.~\ref{Fig_Supp_2} shows nuclear polarization distribution via DNP for an ensemble of 1000 nuclei. Electrons with a zero average spin result in a zero nuclear polarization and  spin-polarized electrons (as generated in TMDs where we excite a selective valley) result in a non-zero nuclear polarization.

We also emphasize that the Overhauser detuning shift is not phenomenological or imposed. Below we derive an effective Hamiltonian for the electron dynamics which contains the Overhauser shift.
Let us begin with the Liouville equation describing single electron-multi nuclear dynamics. Since the analysis is similar for holes as well, we will just concern ourselves with discussing the case for electrons. 
\begin{align}
\frac{d\rho}{dt} = -i [H,\rho] 
\end{align}
The Hamiltonian consists of the hyperfine coupling (where we ignore the flip-flop terms) 
\begin{align}
H&=\sum\limits_{i=1}^N {A_e s_z \sigma_z^i}
\end{align}
where $s_z$ is the electron spin and $\sigma_z^i$ is the $i^{th}$ nuclear spin. When written in the basis $(\uparrow\uparrow,\uparrow\downarrow,\downarrow\uparrow,\downarrow\downarrow)$ (where the first arrow represents electron spin and the second arrow represents a single nuclear spin), the Liouville equation for a single electron-single nucleus becomes 
\begin{align}
\frac{d\rho}{dt} = \left( \begin{array}{cccc}
0& -iA^e\rho_{12} & -iA^e\rho_{13} &0  \\
iA^e\rho_{21} & 0 & 0 & iA^e\rho_{24}  \\
iA^e\rho_{31} & 0 &0 & iA^e\rho_{34}  \\
0 &-iA^e\rho_{42} &-iA^e\rho_{43} &0   \\
\end{array} \right) 
\end{align} 
The Liouville equation for the full system with $N$ nuclei will be just the above matrix forming block-diagonal entries of a $4N\times 4N$ matrix on the R.H.S. Assuming that the electron-nuclear entanglement decays rapidly the full density matrix is the tensor product of an electron part ($\rho_e$) and a nuclear part ($\rho^i_n$), and we can perform a partial trace over nuclear degrees of freedom since the nuclear dynamics is slower compared to the electron dynamics. Therefore the rate equation becomes
\begin{align}
\frac{d\rho_e}{dt} &=  \left( \begin{array}{cc}
0& -iA^e\rho^{12}_e \sum\limits_i(\rho_n^{i\uparrow}-\rho_n^{i\downarrow})  \\
iA^e\rho^{21}_e\sum\limits_i(\rho_n^{i\uparrow}-\rho_n^{i\downarrow}) & 0  \\
\end{array} \right) \\
&=  \left( \begin{array}{cc}
0& -iA^e\rho^{12}_e m  \\
iA^e\rho^{21}_em & 0  \\
\end{array} \right)  \\
&= -i[H_{\text{eff}},\rho_e]
\end{align}
where $m$ is the total nuclear magnetization, and $H_{\text{eff}}=mA_e\sigma_z$ accounts for the Zeeman shift due to dynamic nuclear polarization.

\section{Physical picture-validity of timescale hierarchy}
We consider the hyperfine interaction of one nucleus with one electron for simplicity. The laser excites spin-up electrons in the vicinity of $\mathbf{K}$ valley. The recombination process collapses the electron spin state into the spin-up state  with a high probability on a timescale relevant to the e-h recombination process ($\gamma_1$ below). The collapse on to the spin-down state happens with low probability as determined by the hyperfine interaction timescale. In that case the nuclear spin must flip. We model this physical process by different Kraus operators on the spin-up and spin-down electron state. For the spin-up electron state, the Kraus operator relaxes it to a hole state on a faster timescale ($\gamma_1$ below). The Kraus operator for the spin-down electron state models decoherence which happens at a slower timescale ($\gamma_2$ below). Note that the spin-down state cannot recombine with the hole. The Hamiltonian which describes free evolution via the hyperfine interaction can be written as

\begin{figure}
\includegraphics[scale=0.35]{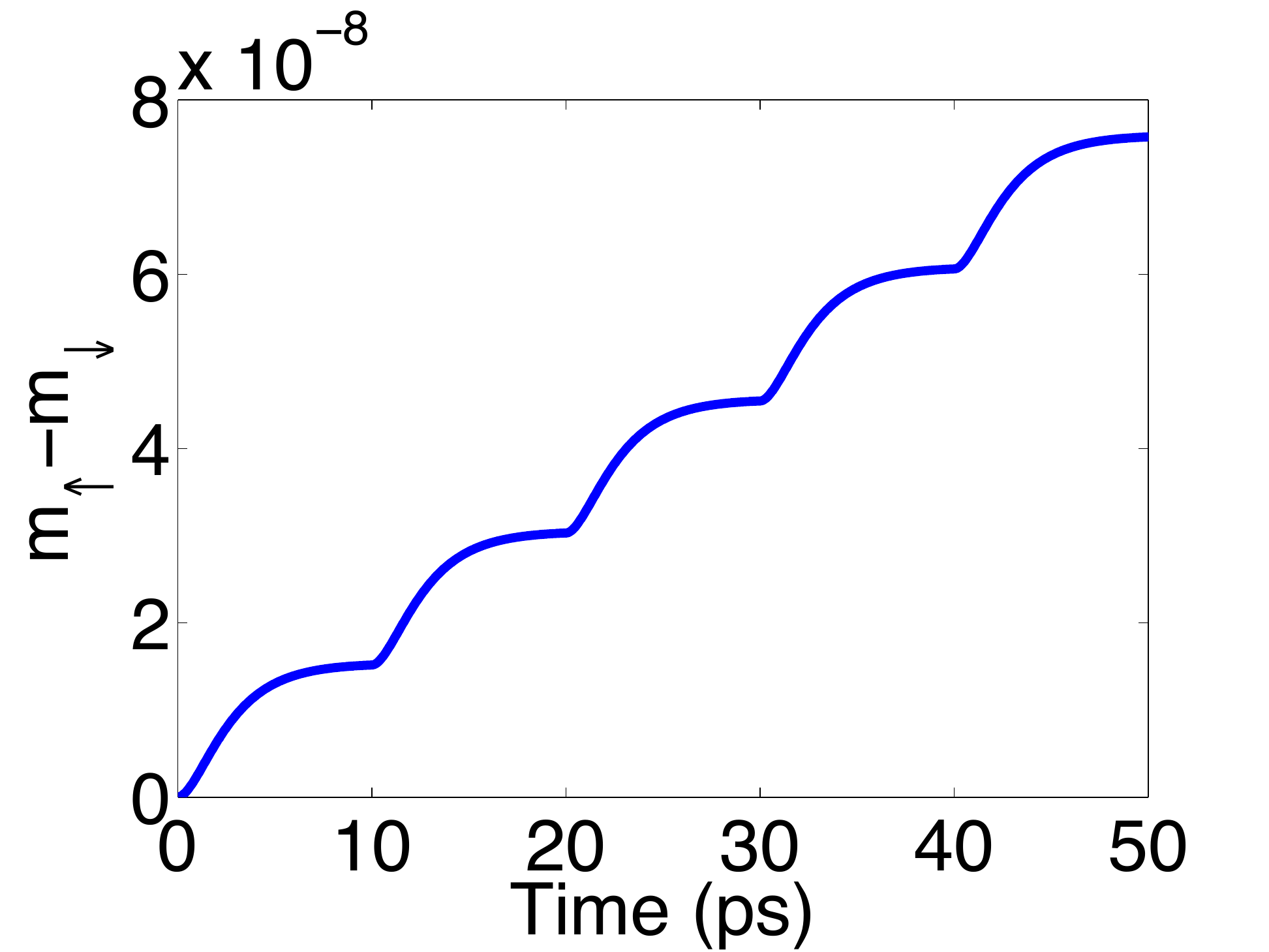}
\caption{The difference between spin-up and spin-down nuclear polarization for a single nucleus as a function of time. Here the first electron (which is spin-up at $t=0$) relaxes (i.e. recombines with the hole) on a timescale of about 10 ps, but shifts the nuclear polarization away from zero by a small amount in that period. Once the electron has completely relaxed, we then repeat the procedure with a new spin-up electron after 10ps. There is a buildup of nuclear polarization demonstrating the validity of timescale hierarchy.}
\label{Fig_Supp_3}
\end{figure} 

\begin{eqnarray}
H = A^e \begin{pmatrix}
  1 & 0 &0 &0 &0 &0 \\
  0 & -1 &\frac{\eta}{2} &0 &0 &0 \\
  0 & \frac{\eta}{2} &-1 &0 &0 &0 \\
  0 & 0 &0 &1 &0 &0 \\
  0 & 0 &0 &0 &0 &0 \\
  0 & 0 &0 &0 &0 &0 
 \end{pmatrix}
\end{eqnarray}
written in the basis $|e^\uparrow n^\uparrow, e^\uparrow n^\downarrow, e^\downarrow n^\uparrow, e^\downarrow n^\downarrow, h^\uparrow n^\uparrow, h^\uparrow n^\downarrow \rangle$, where $e$, $n$, and $h$ represent electron, nucleus, and hole respectively. The arrows represent spin-up or spin-down state. The non-trivial elements of the Lindblad operators ($L_1$ to $L_6$) corresponding to the various Kraus operators are $ \langle h^\uparrow n^\downarrow  |L_1| e^\uparrow n^\downarrow \rangle=\sqrt{\gamma_1}$, $ \langle h^\uparrow n^\uparrow  |L_2| e^\uparrow n^\uparrow \rangle=\sqrt{\gamma_1}$, $\langle e^\uparrow n^\uparrow  |L_3| e^\uparrow n^\uparrow \rangle=\sqrt{\gamma_2}$, $\langle e^\downarrow n^\uparrow  |L_4| e^\downarrow n^\uparrow \rangle=\sqrt{\gamma_2}$, $\langle e^\uparrow n^\downarrow  |L_5| e^\uparrow n^\downarrow \rangle=\sqrt{\gamma_2}$,  $\langle e^\downarrow n^\downarrow  |L_6| e^\downarrow n^\downarrow \rangle=\sqrt{\gamma_2}$.   

We solve for the dynamics via the equation $ \dot{\rho} = -i[H,\rho] +  L[\rho]$, where $L[\rho]$ is constructed via Eq.~\ref{Eq_Supp_Lindblad}. In Fig.~\ref{Fig_Supp_3} we plot the difference between spin-up and spin-down nuclear polarization for a single nucleus as a function of time. We evolve the first electron via the hyperfine interaction with Lindblad operator taking into account relaxation terms. The first electron, which is spin-up at $t=0$ relaxes (i.e. recombines with the hole) on a timescale of about 10 ps, but shifts the nuclear polarization away from zero by a small amount in that period. Once the electron has completely relaxed, we then repeat the procedure with a new spin-up electron after 10ps. The buildup of nuclear polarization is clearly demonstrated which highlights the validity of timescale hierarchy.

\section{Including spin splitting in the conduction band}
\begin{figure}
\includegraphics[scale=0.23]{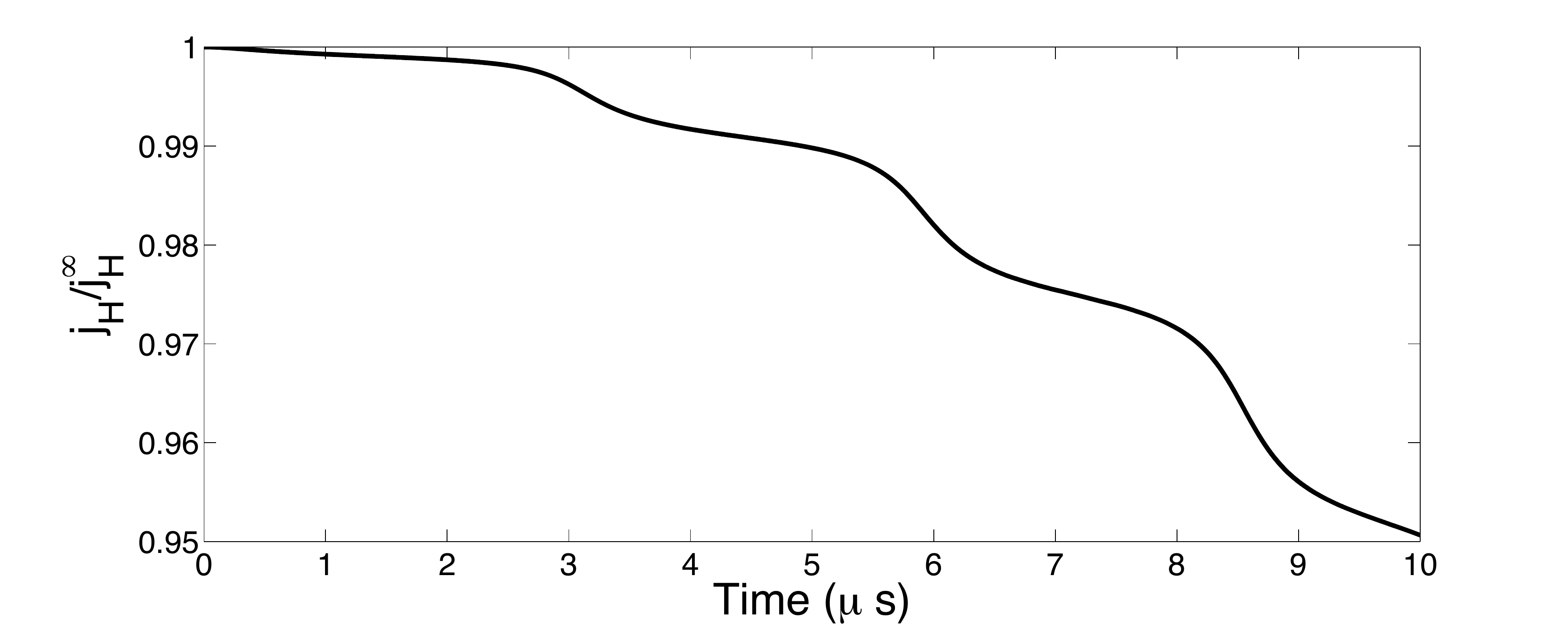}
\includegraphics[scale=0.23]{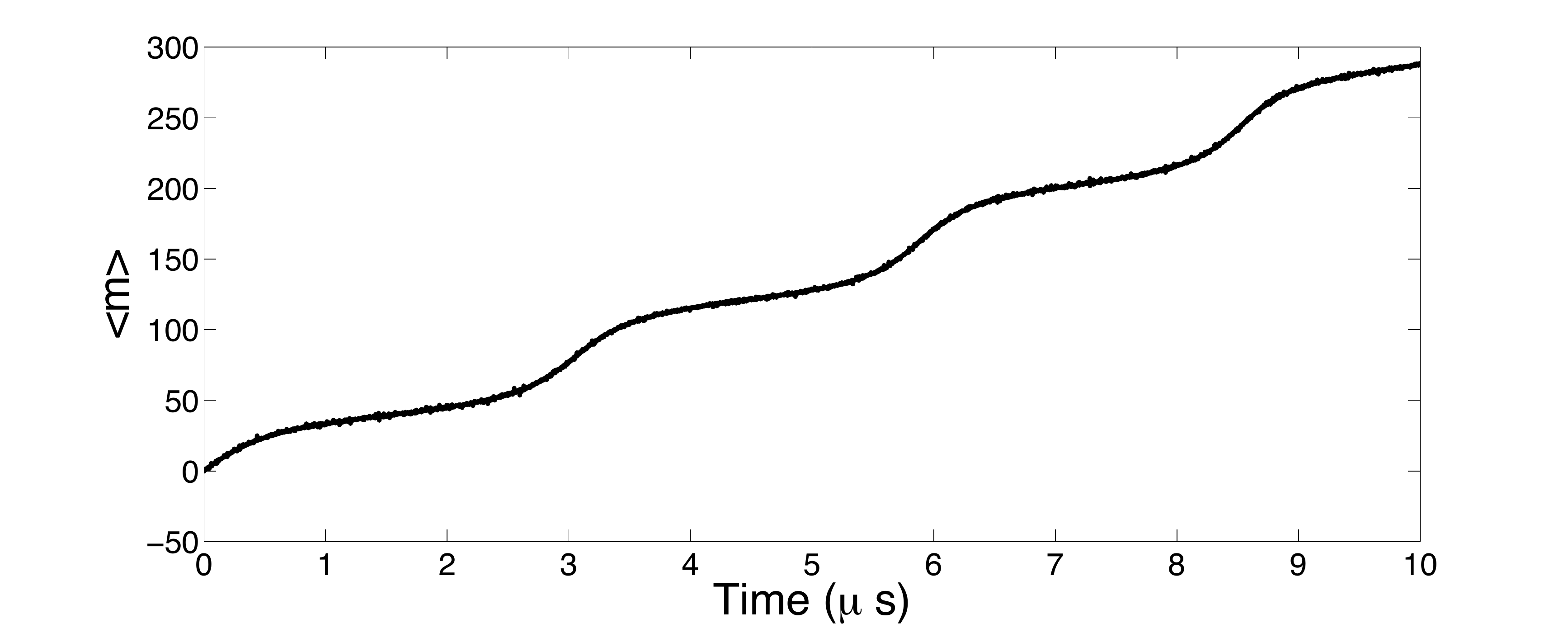}
\caption{Hall current (top) and nuclear magnetization (bottom) for a constant $\sigma^+$ drive. These plots are for the case of $N=10^6$ nuclei, and $\Omega_{R\mathbf{K}}=0.5meV$, including a spin splitting in the conduction band of 5T. The hyperfine values were chosen to be the same as that in Fig.~\ref{Fig_3}.}
\label{Fig_supp_new1}
\end{figure}
The spin orbit coupling in the conduction band is typically a few meV corresponding to a few Teslas. In this section we provide a plot for Hall current and DNP including a splitting in the conduction band. Indeed a similar behavior is seen as seen in Fig. 4 of the main text. Therefore including this spin-splitting does not result in qualitative changes. Fig.~\ref{Fig_supp_new1} shows the Hall current and nuclear magnetization for a constant $\sigma^+$ drive including a conduction band spin splitting of 5T.



\end{document}